\documentclass[a4paper,11pt]{article}
\usepackage{jheppub} 
\usepackage{lineno}
\usepackage{appendix}
\usepackage{graphicx} 
\usepackage{bm}
\usepackage{braket}
\usepackage{amsmath}
\usepackage{amsfonts}
\usepackage{mathtools}
\usepackage{graphicx}
\usepackage{cancel}
\usepackage{slashed}
\usepackage{wrapfig}
\usepackage{nextpage}
\usepackage{enumitem}
\usepackage{float}
\usepackage{hyperref}
\usepackage{amssymb}
\usepackage{enumitem}
\usepackage{simpler-wick}
\hypersetup{colorlinks=true, urlcolor=black, citecolor=blue, linkcolor=black}
\newcommand{\sgn}{\mathrm{sgn}}

\title{\boldmath A Soft Theorem from vertex-like operators in BFSS Theory}

\author[1]{Davide Laurenzano\note{Corresponding author.}}
\author[]{and John F. Wheater}

\affiliation[]{Rudolf Peierls Centre for Theoretical Physics, University of Oxford \\ Parks Road, Oxford OX1 3PU, United Kingdom}

\emailAdd{davide.laurenzano@physics.ox.ac.uk}
\emailAdd{john.wheater@physics.ox.ac.uk}

\abstract{In this paper, we derive a soft theorem at leading and subleading orders within the context of BFSS matrix theory. Specifically, we consider the effective field theory describing interactions between bound states of $D_0$-branes at leading order, which are dual to supergraviton interactions in the eleven-dimensional target space. This theory is obtained from BFSS theory by integrating out heavy degrees of freedom in the large-distance limit at one loop. As part of our analysis, we demonstrate that when treated as a one-dimensional quantum field theory with a UV cutoff, the theory is super-renormalizable and all Feynman diagrams converge. Our main result shows that the theory admits vertex-like operators with the correct quantum numbers to represent supergravitons in target space and that their correlation functions exhibit soft factorisation at both leading and subleading orders.}

\begin{document}
\maketitle 
\flushbottom
\date{}

\section{Introduction}
Recent advances in celestial holography 
\cite{Strominger_2014,He_2016, Kapec_2017, Kapec_2022, He_2014, He_2015, Kapec_2014, Campiglia_2014,cachazo2014evidence, campiglia2015asymptotic,  Colferai_2021, Agrawal:2025}
have unveiled a fundamental link between soft theorems \cite{Weinberg} and asymptotic symmetries. Notably, this connection has been extended in \cite{IR,miller2022soft} to the BFSS matrix theory, also known as M(atrix) Theory, which is a conjectured dual description of M-Theory in the Discrete Light Cone Quantisation (DLCQ) framework \cite{BFSS, susskind1997conjecture}. 
BFSS theory is a matrix quantum mechanics theory describing the non-relativistic limit of $D_0$ brane dynamics. It is obtained as the dimensional reduction, from $d=10$ to $d=1$ (i.e. only time remains) of a $SU(N)$ supersymmetric Yang-Mills theory with one supercharge. The resulting field content is a nine-component vector field $X^{I}$, $I=1,...,9$ and a 16-component spinor $\psi_{\alpha}$, both valued in the $su(N)$ adjoint representation. Here $I$ and $\alpha$ are target space indices. 
The Hamiltonian of the theory is
\begin{equation}
    H=R \, Tr\biggl[\frac{1}{2}P^{I}P_{I}+\frac{1}{4}[X^{I},X^{I}]^2+\psi^t_{\alpha}\Gamma^{\alpha \beta}_{I}[\psi_{\beta}, X^{I}]\biggr] \, ,
    \label{eq:HBFSS}
\end{equation}
where $P^{I}$ is the canonical conjugate momentum to $X^{I}$ and $\Gamma_{I}$ are the $spin(9)$ gamma matrices.

The BFSS conjecture \cite{BFSS, susskind1997conjecture} states that this theory is in fact dual to M-theory in Discrete Light Cone Quantisation (DLCQ) (see \cite{Taylor_2001, Polchinski_1999, ydri2018review, bigatti1997review} for reviews). A number of consistency checks for this conjecture have been given \cite{Seiberg_1997, Sethi_1998, Yi_1997, Bergner_2021,Pateloudis_2022}. 
DLCQ is a compactification scheme from eleven to ten dimensions. The compactified variable is $x^-=\frac{x^0-x^{10}}{\sqrt{2}}\sim x^-+2\pi R$, $R$ being the radius of compactification which plays the role of coupling constant in the matrix model dual (\ref{eq:HBFSS}).  As a consequence, the $p^+=\frac{p^0+p^{10}}{\sqrt{2}}$ component of the momentum is quantized by
\begin{equation}
    p^+=\frac{N}{R} \,,
\end{equation}
and the compactification quantum number $N$ is mapped, in the dual description, into the matrix size. In lightcone coordinates, $x^+$ plays the role of time, and the conjugate momentum, $p^-$, is then the energy. The M-theory limit consists of taking the radius of compactification to infinity, in such a way that momentum components are kept finite. Then, in order to keep $p^+$ finite, the scaling must be
\begin{equation}
    N \rightarrow \infty\,, \qquad R = R_0 \times N\,,
    \label{eq:scaling}
\end{equation}
where $R_0$ is a dimensionful constant. 
The degrees of freedom of Matrix Theory include $D_0$ brane bound states \cite{DANIELSSON_1996, Kabat_1996,Bachas_1996,Moore_2000}, which have the same quantum numbers as the supergraviton multiplet in the low energy description of M-theory, namely supergravity. 
Most of the consistency checks of the BFSS conjecture that have been performed over the years have revolved around checking the agreement between scattering amplitudes of (super)gravitons in supergravity and those of $D_0$ particles in Matrix Theory. The matching of the corresponding effective interaction terms in the two theories played a pivotal role \cite{Okawa_1999, Okawa_1999_2, Becker_1997, Becker_1997_2,Kabat_1998, IV_1999, Plefka_1998, Plefka2_1998, Serone_2000, plefka1998asymptotic, Helling_1999, Becker_1998} (see  \cite{Lin:2024, Biggs:2025} for recent developments in perturbative computations). However, there is no conclusive proof that Matrix Theory reproduces the full 11-dimensional Lorentz group.

In two recent papers \cite{IR,miller2022soft}, it was shown that the validity of a soft theorem for BFSS amplitudes would imply the conservation of infinitely many charges associated with asymptotic symmetries, which include and generalise the Poincaré group. This development has enabled a deeper comprehension of how the complete 11-dimensional Lorentz group manifests itself within the BFSS theory, alongside the recognition of an infinite-dimensional asymptotic symmetry group.
Soft theorems \cite{Weinberg} hold for a vast class of theories describing massless particles (including supergravity); they relate the scattering amplitude of states which include a soft particle with the amplitude between the corresponding states when such a particle is removed. However, it was not until recently that it was realised that soft theorems are equivalent to Ward identities for the conservation of charges associated with asymptotic symmetries \cite{Strominger_2014,He_2016, Kapec_2017, Kapec_2022, He_2014, He_2015, Kapec_2014, Campiglia_2014,cachazo2014evidence, campiglia2015asymptotic,  Colferai_2021} (c.f. \cite{McLoughlin_2022, Donnay:2023} for recent reviews of asymptotic symmetries in celestial holography). These are diffeomorphisms of asymptotically flat space-time generated by vector fields which become Killing vectors in the asymptotic region. The statement of \cite{IR, miller2022soft} is that indeed this connection is also present in the BFSS theory. Soft theorems have been shown to be universally valid \cite{Broedel_2014, Bern_2014,Bianchi_2015, Di_Vecchia_2015, Sen_2017, Higuchi_2018, strominger2018lectures}, therefore it would be natural to expect a soft theorem to hold in BFSS if the latter were to describe supergraviton dynamics. 

A proof of such a theorem was given in 
\cite{Herderschee_2023, Maldacena_soft}. 
That proof was carried out by first showing that any scattering amplitude involving massless particles and satisfying a specific set of assumptions enjoys a soft factorisation. It was then argued that those assumptions should be satisfied by Matrix Theory scattering amplitudes, under the condition that amplitudes involving five or more non-soft particles are evaluated in a particular kinematic regime.

In this paper, we approach the problem from a complementary perspective by proving a soft theorem for the effective theory describing supergraviton interactions in Matrix Theory. Specifically, we will show that certain composite operators, possessing the right quantum numbers to represent gravitons in the target space, have correlation functions that enjoy a soft factorisation to the leading and subleading orders in the large $N$ limit. 

The paper is organized as follows: in section \ref{sec:theorem} we analyze the statement of the soft theorem for BFSS theory; in section \ref{sec:setup} we outline the formalism in which the computation is carried out and the strategy we will adopt; section \ref{sec:leading} will then be devoted to the detailed proof of the leading soft theorem, while in section \ref{sec:subleading} we will obtain the subleading soft factor; in section \ref{sec:conclusion} we discuss the result and provide an outlook on future developments of this work.
\section{The Theorem}
\label{sec:theorem}
The soft graviton theorem provides a relationship between scattering amplitudes of the form (we will use the notation of \cite{IR})
 \begin{equation}
\mathcal{A}(q_s, h_s;p_1, \theta_1;...; p_n, \theta_n)=
(S^{(-1)}+S^{(0)}+...)\mathcal{A}(p_1, \theta_1;...; p_n, \theta_n) \, ,
\label{eq:softtheoremgeneral}
\end{equation}
where momenta are characterized by an overall scale $|p^{\mu}|\propto \omega$, and $h_s$, $\theta_i$ are graviton polarizations. The particle with momentum $q_s$ and polarization $h_s$ is soft, namely $\omega_s \ll \omega_i, \forall i$, and the soft factors are such that 
\begin{equation*}
    S^{(n)}\sim \Bigl(\frac{\omega_s}{\omega_i}\Bigr)^n \,.
\end{equation*}
In particular, the leading and subleading factors take the form
\begin{equation}
    S^{(-1)}=\frac{k}{2}h_{s\mu \nu}\sum_{j=1}^{n} \eta_s \eta_j\frac{p_j^{\mu}p_j^{\nu}}{q_s \cdot p_j}, \qquad S^{(0)}=i\frac{k}{2}h_{s\mu \nu} \sum_{j=1}^n\eta_s \eta_j \frac{p_j^{\mu} J_j^{\nu \rho}q_{s,\rho}}{q_s \cdot p_j}\, ,
    \label{eq:leadsub}
\end{equation}
where $\eta=-1$ if the corresponding particle is \textit{incoming} and $\eta=1$ if \textit{outgoing}, $J_j^{\mu \nu}=L_j^{\mu \nu}+S_j^{\mu \nu}$ stands for the total angular momentum of the $j$th-particle and $k=\sqrt{32\pi G_N}$.
Soft terms are ubiquitous in theories with massless particles.
We are concerned here with the case of 11-dimensional supergravity, so $\mu=0,1...,10$.

To prove a result analogous to (\ref{eq:softtheoremgeneral}) for BFSS theory, we will use the large distance effective theory describing interactions of $D_0$ brane bound states. 
This effective (quantum mechanical) interaction between two bound states is obtained \cite{BFSS, Kabat_1998, IV_1999} by integrating out quantum fluctuations $Y$ around a classical background
\begin{equation}
    X^I_{back}=X_1^I \mathbb{I}_{N_1} \oplus X_2^I \mathbb{I}_{N_2}, \quad X^I=X^I_{back}+Y^I.
\end{equation}
Then, by defining $r$ as the center of mass distance between particles at some initial reference time, we can interpret, for large distances, the background matrix as describing two $D_0$ brane bound states composed of $N_i$ branes and having transverse coordinates $X^I_i$. Quantum fluctuations give rise to massive excitations, both bosonic and fermionic, whose mass squared is proportional to the distance squared, $M^2_{B/F} \propto r^2$. By integrating out these modes at large distances, we obtain the effective theory for $D_0$ bound state scattering
\begin{equation}
    \mathcal{L}=\frac{N_1}{2R}{(\Dot{X}^I_1)^2}+\frac{N_2}{2R}{(\Dot{X}^I_2)^2}-A\frac{N_1 N_2}{R^3 r^7}\bigl[(\Dot{X}_1^{I}-\Dot{X}_2^{I})^2\bigr]^2+\mathcal{L}_{ang}+\mathcal{L}_{spin}\, , 
    \label{eq:effective}
\end{equation}
where $A$ is an order $1$ constant (in units where the Planck length is set to $1$). The first interaction term will be sufficient for the computation of $S^{(-1)}$, while the derivation of $S^{(0)}$ will require the explicit expression, which we will provide later, of the subleading contributions $\mathcal{L}_{ang}$ and $\mathcal{L}_{spin}$.
The Lagrangian (\ref{eq:effective}) can be naturally generalized to describe $n$ such particles with pairwise interaction, leading to
\begin{equation}
    \mathcal{L}_n=\sum_{i=1}^n \frac{N_i}{2R} (\Dot{X}^I_i)^2 -\sum_{\substack{i \ne j \\ i,j=1}}^n A\frac{N_i N_j}{R^3 r^7}\bigl[(\Dot{X}_i^{I}-\Dot{X}_j^{I})^2\bigr]^2+\mathcal{L}_{ang}+\mathcal{L}_{spin}. \label{eq:effectiven}
\end{equation}
In this approximation, we have ignored three-body interactions, which will appear at higher orders in the effective field theory. At large distances, where $in$ and $out$ particles are non-interacting, their non-relativistic momentum will be given by $p_i^I=\frac{N_i}{R}v^I_i$, where $v^I_i$ is the velocity eigenvalue of $\Dot{X}_i^I$. 

We define asymptotic \textit{in} and \textit{out} states (corresponding to Matrix Theory time $\mp\infty$ respectively) as multi-$D_0$ brane bound states, each being characterised by the number of particles in the bound state, the nine-dimensional velocity vector and the $SO(9)$ polarisation. An $n$ particle \textit{in} state is then written as
\begin{equation}
    \ket{N_1,v^{I}_1,\theta_1; ...; N_n, v^{I}_n, \theta_n}_\text{in}, \quad N_1+N_2+...+N_n=N \, ,
\end{equation}
and analogously for \textit{out} states. Consequently, the scattering amplitude between $n$ multi-particle states is given by the $S$-matrix element 
\begin{equation*}
        \mathcal{A}(N_1, v^{I}_1, \theta_1;...; N_n, v^{I}_n, \theta_n)
        =\prescript{}{\text{out}}{}\braket{N_1, v^{I}_1, \theta_1;...; N_j,v^{I}_j, \theta_j|N_{j+1}, v^{I}_{j+1}, \theta_{j+1};...;N_n, v^{I}_n, \theta_n}_\text{in}.
\end{equation*}
In this model, the parameter characterising the soft expansion is the number of branes in the bound state. In particular, a particle is soft (and its quantum numbers will be labelled by the subscript `$s$') if $\frac{N_s}{N_j} \ll 1 \, \, \forall j$, so that the number of $D_0$ branes in the bound state is dwarfed by that of any other state. Then, the soft theorem can be stated as
 \begin{multline}
\mathcal{A}(N_s, v^{I}_s, h_s;N_1, v^{I}_1, \theta_1;...; N_n, v^{I}_n, \theta_n)=
\\
=(S^{(-1)}+S^{(0)}+...)\mathcal{A}(N_1, v^{I}_1, \theta_1;...; N_n, v^{I}_n, \theta_n) \, ,
\label{eq:softtheorem}
\end{multline}
where the expansion in soft factors is such that $S^{(k)} \sim \Bigl(\frac{N_s}{N_j}\Bigr)^k$. In the large $N$ limit of Matrix Theory, the soft limit corresponds to the scaling $N_s \sim O(1)$, $N_j \sim N \rightarrow \infty$. Then the soft factors scale according to $S^{(k)} \sim \Bigl(\frac{1}{N}\Bigr)^k$.

The explicit form for $S^{(-1)}$ and $S^{(0)}$, dubbed leading and subleading respectively, in BFSS has been computed in \cite{IR}. The expressions in (\ref{eq:leadsub}) are translated to the $D_0$ brane language by setting
\begin{equation}
p^{\mu}=\frac{N}{\sqrt{2}R}(1+v^2, 2v^I,1-v^2) \, ,
\label{eq:dictionary}
\end{equation} 
where $v^I$ is the 9-dimensional transverse velocity and $v^2=v^I v_I$. The polarisation tensor $h_{\mu\nu}$ is described by the symmetric traceless SO(9) tensor $h_{IJ}$, which describes the 44-dimensional polarisation basis of 11-dimensional gravity. We find
\begin{equation}
    S^{(-1)}=-2k\,h_{IJ}\sum_j\eta_s \eta_j\frac{N_j}{N_s}\frac{v^I_{sj} v^J_{sj}}{v^2_{sj}} \,,
\end{equation}
and the subleading term is given later in \eqref{eqn:subleadingS}.
In \cite{IR} it was shown that $S^{(-1)}$ and $S^{(0)}$ are associated with the conservation of BFSS charges and the consequent existence of asymptotic symmetry groups.
 Our goal will be to demonstrate the validity of (\ref{eq:softtheorem}) within the effective field theory \eqref{eq:effectiven}.
\section{The Formalism}
\label{sec:setup}
In this section, we outline the formalism we will use to give a world-line proof for the soft theorem. 
We consider the one-dimensional quantum field theory of fields $X^I_i(t)$, $i=1\ldots n$, defined by a path integral with Lagrangian given by (\ref{eq:effectiven}) and compute scattering amplitudes in a vertex operator formalism, which naturally accounts for the translation between world-line and target space quantities. In particular, the scattering amplitude involving $n$ asymptotic particles takes the form 
\begin{align}
    \mathcal{A}(N_s, v^{I}_s, h_s;N_1, v^{I}_1, \theta_1;...; N_n, v^{I}_n, \theta_n)=\prod_{i=1}^n \int d\mu_{t_i}\braket{\Omega|T[\mathcal{V}^{*}_1(t_1)...\mathcal{V}_n(t_n)]|\Omega} \nonumber
    \\
   =\prod_{i=1}^n \int d\mu_{t_i}\braket{\mathcal{V}^{*}_1(t_1)...\mathcal{V}_n(t_n) } =\prod_{i=1}^n \int d\mu_{t_i}\frac{\int \Pi_{k=1}^n \mathcal{D} X_k \,\mathcal{V}_k(t_k)e^{iS}}{\int \Pi_{k=1}^n \mathcal{D} X_k \,e^{iS}} \, ,
   \label{eq:Green}
\end{align}   
where $\braket{}$ denotes the expectation value with respect to the full action and $T$ stands for time ordering. Moreover, we define
\begin{equation}
    \int d\mu_{t_i}\;:= \lim_{\alpha \rightarrow  0}\int_{-\infty}^{+\infty} dt_i e^{-\alpha |t_i|}\;,
\end{equation}
namely, we integrate over operator insertion time, with an exponential regulator that kills divergences for large times. In the particular case of soft particle emission, the amplitude is
\begin{equation}
    \mathcal{A}(N_s, v^{I}_s, h_s;N_1, v^{I}_1, \theta_1;...; N_n, v^{I}_n, \theta_n)=\int d\mu_{t_s}\prod_{i=1}^n \int d\mu_{t_i}\braket{\Omega|T[\mathcal{V}^{*}_s(t_s)\mathcal{V}^{*}_1(t_1)...\mathcal{V}_n(t_n)]|\Omega} \, .
    \label{eq:softAmp}
\end{equation}
The operators appearing in the expressions (\ref{eq:Green}) and (\ref{eq:softAmp}) are vertex-like operators and represent asymptotic states in the target space. In particular, starred vertex operators create particles in the \emph{out} state, while the non-starred create them in the \emph{in} state. We describe their properties in subsection \ref{sec:vertop}. First, we will outline the formalism in which we will compute (\ref{eq:softAmp}). 
\subsection{Correlation function in the effective theory}
\begin{figure}[h]
\centering
\includegraphics[width=0.4\textwidth]{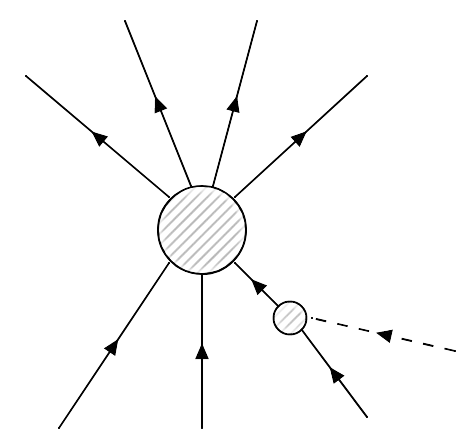}
\caption{Soft absorption. The dashed line represents a soft $D_0$ brane bound state, while the solid lines represent the hard ones. The big blob represents interactions between hard particles, whereas the small one stands for interactions between a soft and a hard particle.}
\label{fig:absorbtion}
\end{figure}

We evaluate \eqref{eq:softAmp} using perturbation theory in the interaction term of (\ref{eq:effective}). The amplitude (\ref{eq:Green}) represents a correlation function in the 11-dimensional target space. 
It follows that the target space picture of (\ref{eq:softAmp}) is that of a soft emission or absorption, in which the soft $D_0$ brane bound state is attached to one of the hard external legs, see figure \ref{fig:absorbtion}.
We therefore consider the series expansion corresponding to the interaction of the soft $D_0$ brane bound state with one of the hard particles, to all orders in perturbation theory. As we will see in the next subsection, each vertex operator is a composite operator depending on $X_j(t)$ fields and creates a particle in the asymptotic initial or final state, so the interaction time runs up to the earliest insertion time of the two particles involved in the case of \textit{outgoing} particles, or from the latest insertion time in the case of \textit{incoming} particles, or between the two in case of one \textit{incoming} and one \textit{outgoing} particle. In the following, we can pick the case in which both particles are in the \emph{out} state for convenience, but the results can be easily extended to the other cases.

Defining
\begin{equation}
     S_{s j}=\sum_{n=0}^{\infty}
     \int_{-\infty}^{\min(t_j,t_s)}du_1...du_n\frac{i^n}{n!}\mathcal{L}_{s j}(u_1)...\mathcal{L}_{s j}(u_n)
     \, ,
     \label{eq:dysonseries}
\end{equation}
where $\mathcal{L}_{s j}$ is the interaction term in (\ref{eq:effective}) with one of the interacting particles being soft, the correlation function \eqref{eq:softAmp} can be written as
\begin{multline}
    \mathcal{A}(N_s, v^{I}_s, h_s;N_1, v^{I}_1, \theta_1;...; N_n, v^{I}_n, \theta_n) \\
   = \int d\mu_{t_s}\prod_{i=1}^n \int d\mu_{t_i}\sum_{j=1}^{n}\braket{\mathcal{V}^{*}_1(t_1)...\overline{\mathcal{V}^{*}_s(t_s)S_{s j}\mathcal{V}^{*}_j(t_j)}...\mathcal{V}_n(t_n)}_h \, .
    \label{eq:ampint}
\end{multline}
Here, the expectation value $\braket{}_h$ is defined with respect to an action containing only hard fields $X^I_i(t)$, $i=1\ldots n$; and the overline indicates that the soft field 
$X^I_s(t)$ is contracted out of the string $\mathcal{V}^{*}_s(t_s) S_{sj}(u) \mathcal{V}^{*}_j(t_j)$. We will show how this computation yields a soft factorization which reproduces (\ref{eq:softtheorem}). Note that the effective action (\ref{eq:effective}) describes graviton interactions with no transfer of longitudinal momentum $p^+$. We will be able to use it to characterise soft emission/absorption because the longitudinal momentum transfer is negligible in the soft limit.

Before proceeding, it is helpful to introduce an auxiliary field. (For the moment, we focus on the leading interaction; in Section \ref{sec:subleading} we will extend the discussion to include the subleading contributions $\mathcal{L}_{ang}$ and $\mathcal{L}_{spin}$.)
After a field redefinition, the Lagrangian has a canonically normalised kinetic term
\begin{equation}
    \mathcal{L}=\frac{1}{2}(\Dot{X}^I_s)^2+\frac{1}{2}(\Dot{X}^I_j)^2-\frac{\lambda}{4}\biggl[\Bigl(\sqrt{\frac{R}{N_s}}\Dot{X}^I_s-\sqrt{\frac{R}{N_j}}\Dot{X}^I_j\Bigr)^2\biggr]^2 \, ,
\end{equation}
where $\lambda=\frac{4 AN_s N_j}{R^3 r^7}$.
Then we introduce an auxiliary field $\sigma(t)$ and write the partition function as 
\begin{equation}
    Z=const \int \mathcal{D}X_s \mathcal{D}X_j \mathcal{D}\sigma \,e^{i\int dt \mathcal{L}'} \, ,
\end{equation}
where
\begin{equation}
    \mathcal{L}'=\frac{1}{2}(\Dot{X}^I_s)^2+\frac{1}{2}(\Dot{X}^I_j)^2+\frac{\lambda}{4}\sigma^2-\frac{\lambda}{2}\sigma \Bigl(\sqrt{\frac{R}{N_s}}\Dot{X}_s^I-\sqrt{\frac{R}{N_j}}\Dot{X}_j^I\Bigr)^2.
    \label{eq:Lagaux}
\end{equation}
The quadratic Yukawa-like interaction is more convenient to manipulate. 

Standard graph combinatorial techniques show that the superficial degree of divergence of diagrams scales in the same way as powers of the coupling constant. This suggests that the expansion parameter for the perturbative computation is $\frac{\lambda}{\epsilon}$, where $\epsilon$ is a short distance cutoff which must be sent to zero at the end of the computation. To implement this cutoff, we introduce a kinetic term for $\sigma$
\begin{equation}
        \mathcal{L}^{kin}_{\sigma}=\frac{\epsilon^2 \lambda}{4} \Dot{\sigma}^2 \, .
        \label{eq:kinreg}
\end{equation}
Note that, as explained around equation (\ref{eq:scaling}), in DLCQ, the $p^+$ component of the momentum is kept fixed in the large $N$ limit. As a consequence, the light-like radius scales like $R\sim N$ and 
\begin{equation}
    \lambda=\frac{4AN_s N_j}{R^3 r^7}
    \label{eq:coupling}
\end{equation} scales like $\sim \frac{1}{N^2}$ in the soft limit where $N_j \sim N$ and $N_s \sim O(1)$. In fact, as discussed in \cite{Polchinski_1999}, the flat space M-Theory regime corresponds to $N^{\frac{1}{9}}<r<N^{\frac{1}{7}}$ in Planck units. Therefore, in this region of parameter space, $\lambda$ decays faster than $\frac{1}{N^2}$. Moreover, $\lambda$ depends on time through the $\frac{1}{r^7}$ factor; for simplicity of notation, we will only write down this dependence explicitly when relevant. The Lagrangian of the theory now takes the form 
\begin{multline}
    \mathcal{L}_{ren}=\mathcal{L}'+\mathcal{L}^{kin}_{\sigma}\\
    =\frac{1}{2}(\dot{X}^I_s)^2+\frac{1}{2}(\dot{X}^I_j)^2+\frac{\epsilon^{2} \lambda}{4}\dot{\sigma}^2+\frac{ \lambda}{4}\sigma^2-\frac{\lambda}{2}\sigma \Bigl(\sqrt{\frac{R}{N_s}}\dot{X}_s^I-\sqrt{\frac{R}{N_j}}\dot{X}_j^I\Bigr)^2 \, ;
    \label{eq:finallag}
\end{multline}
 note that, after introducing the $\epsilon$-regulator through the kinetic term (\ref{eq:kinreg}), the theory is UV finite, and does not require counterterms. This is a special feature of this one-dimensional quantum field theory. We provide a proof of this statement in appendix \ref{sec:app}. Note that we can remove the cutoff after taking the $N \rightarrow \infty$ limit. 
 Indeed, the scattering amplitude still takes the form (\ref{eq:ampint}), but now the interaction term in the Dyson series amounts to
\begin{equation}
    \mathcal{L}_{s j}=-\frac{\lambda}{2} \sigma\Bigl(\sqrt{\frac{R}{N_s}}\dot{X}_s^I-\sqrt{\frac{R}{N_j}}\dot{X}_j^I\Bigr)^2.
    \label{eq:leadingint}
\end{equation}

In the following, we will need the contractions between elementary operators, which can be derived from (\ref{eq:finallag}). Those will be used after we explicitly expand vertex-like operators appearing in (\ref{eq:softAmp}) in terms of $X_j(t)$ fields. First, a contraction between position operators produces a two-point function 
\begin{equation}
    \wick{\c X^{I}_i(t_1) \c X^{J}_j(t_2)}=\braket{T[X^{I}_i(t_1)X^{J}_j(t_2)]}=-\frac{i}{2}\delta_{i j}\delta^{I J} |t_1-t_2|\,,
    \label{eq:pospos}
\end{equation}
as obtained by considering a principal value prescription for the Fourier transform. Consequently, the contraction between a position and a velocity yields
\begin{equation}
    \wick{\c X^{I}_i(t_1) \dot{X}^{J}_j\c(t_2)}=\braket{T[X^{I}_i(t_1)\dot{X}^{J}_j(t_2)]}=\frac{d}{dt_2}\braket{T[X^{I}_i(t_1)X^{J}_j(t_2)]}=\frac{i}{2}\delta_{ij}\delta^{I J} \sgn(t_1-t_2)\,,
\end{equation}
and between two velocities gives 
\begin{equation}
    \wick{\dot{X}^{I}_i\c (t_1) \dot{X}^{J}_j\c (t_2)}=i\delta_{ij}\delta^{I J}\delta(t_1-t_2)\,.
    \label{eq:twovel}
\end{equation}
Using these relations, we obtain
\begin{equation}
    \wick{\c1 e^{i p \cdot X_i(t_1)}\c1 X^{I}_j(t_2)}=\frac{p^{I}}{2}|t_1-t_2|\,e^{i p \cdot X_i(t_1)}\delta_{ij}\,,
\end{equation}
and
\begin{equation}
    \wick{\c1 e^{i p \cdot X_i(t_1)} \dot{X}^{I}_j\c1(t_2)}=-\frac{p^{I}}{2} \,\sgn(t_1-t_2)\,e^{i p \cdot X_i(t_1)}\delta_{ij}\,.
    \label{eq:velexp}
\end{equation}
Finally, the Wick contraction of two $\sigma$ fields is given by 
\begin{equation}
    \wick{\sigma \c (t) \sigma \c (t')}=\braket{\sigma(t)\sigma(t')}=\frac{i}{\epsilon \sqrt{\lambda(t)\lambda(t')}}e^{-\frac{|t-t'|}{\epsilon}}\,.
    \label{eq:twoaux}
\end{equation}
Having outlined the general framework for our computation, we can move on to consider the explicit form of vertex-like operators. 
\subsection{Vertex-like operators}
\label{sec:vertop}
The operators whose correlation functions are considered in the following take the form
\begin{equation}
    \mathcal{V}_j(t_j)= \mathcal{N}_j\,p_j^+:A_{IJ}\Dot{X}^{I}_j(t_j)\Dot{X}^{J}_j(t_j)e^{i X_j(t_j)\cdot p_j }\cos{\Bigl(\beta_j \int_{-\infty}^{+\infty} du \,\dot{X}_j(u)\cdot p_j\Bigr)}: \, ,
    \label{eq:hardvert}
    \end{equation} 
    where $a\cdot b\equiv a_Ib^I$ and $\mathcal{N}_j$ is a normalisation. 
    The operator (\ref{eq:hardvert}) is a vertex-like operator which should describe a massless 11-dimensional particle in the asymptotic region. The cosine term, which deviates from the usual vertex-operator form, is required to reabsorb a divergence in the two-point function, as we will show shortly. This is due to the one-dimensional nature of the theory, which prevents us from having a standard vertex operator structure. Nonetheless, our goal in the rest of the paper will be to show that operators in the form (\ref{eq:hardvert}) reproduce interesting results for target space scattering amplitudes. The transverse momentum is
    \begin{equation*}
        p^{I}_j=\sqrt{2}\frac{N_j}{R}v_j^I\,,
    \end{equation*} 
    which is uplifted to a light-like $11$-dimensional vector by the dictionary (\ref{eq:dictionary}), while
    \begin{equation*}
      p_j^+=\frac{N_j}{R} \, ,
    \end{equation*} 
    and $p_j^-$ is the energy in the target space, such that the on-shell condition 
    \begin{equation*}
        p_j^-=\frac{N_j}{R}{\Vec v_j}^2
    \end{equation*}
   is satisfied.
    The polarisation $A_{IJ}$ is an $SO(9)$ tensor, and is, therefore, a representation of the little group of $11$-dimensional Lorentz symmetry, thus describing a massless particle in $11$-dimensions. We will be interested in the case of a graviton, for which the tensor has rank $2$. The construction can, however, be extended to higher rank representations of $SO(9)$. The structure of (\ref{eq:hardvert}) is determined by the requirement of asymptotic translation and rotation invariance. The normal ordering of terms in the vertex operator will be implied from now on, even when not written explicitly. As a consequence, we will never generate Wick contractions between elements of the same vertex operator. The soft vertex operators correspond to a soft graviton insertion
\begin{equation}
    \mathcal{V}_s(t_s)=\mathcal{N}_s\,q_s^+\,h_{I J} :\dot{X}^{I}_s(t_s)\dot{X}^{J}_s(t_s)
    e^{i q_s \cdot X_s(t_s)}\cos{\Bigl(\beta_s \int_{-\infty}^{+\infty} du \,\dot{X}_s(u)\cdot p_s\Bigr)}: \, .
    \label{eq:softvert}
\end{equation}
Here $q_s$ is a soft momentum (i.e. the overall scale is given by $\frac{N_s}{R} \ll \frac{1}{R_0}$ in the large $N$ limit) and $h_{IJ}$, which is symmetric and traceless, is the graviton polarisation. We perform a field redefinition, such that the kinetic terms in (\ref{eq:effective}) are canonically normalised. Hence, the vertex operator becomes
\begin{equation}
     \mathcal{V}_s(t_s)=\mathcal{N}_s\,h_{I J} :\dot{X}^{I}_s(t_s)\dot{X}^J_s(t_s)e^{i\sqrt{\frac{2N_s}{R}}v_s \cdot X_s(t_s)}\cos{\Bigl(\beta_s\sqrt{\frac{2N_s}{R}} \int_{-\infty}^{+\infty} du \,\dot{X}_s(u)\cdot v_s\Bigr)}: \,.
     \label{eq:vertnorm}
\end{equation}
One of the central claims of the present paper is that operators in the form (\ref{eq:vertnorm}) are suitable to create an asymptotic graviton state. In order to check the statement, we compute the two-point function of vertex operators and check it against the overlap of graviton states. Consider the correlator between two graviton vertex operators in the form (\ref{eq:vertnorm}), with the same number of branes 
\begin{equation}
\mathcal{A}(N_k, v_k, h;N_k, \Tilde{v}_k, \Tilde{h})=\int d\mu_{t_1} d\mu_{t_2}\langle T[\mathcal{V}_k^{*}(t_1)\Tilde{\mathcal{V}}_k(t_2)] \rangle \, .
\end{equation}
Defining 
\begin{equation*}
    T=t_1+t_2, \quad t_k=t_2-t_1
\end{equation*}
we obtain
\begin{equation}
\mathcal{A}(N_k, v_k, h;N_k, \Tilde{v}_k, \Tilde{h})=\int d\mu_{T}\, d\mu_{t_k}\langle T[\mathcal{V}_k^{*}(0)\mathcal{V}_k(t_k)] \rangle
\end{equation}
which is the statement that the expectation value only depends on time differences. 
Analogously, we can make the overall zero-mode of $X_k(t)$ fields explicit, which corresponds to a constant shift 
\begin{equation*}
    X_k^I(t)=\Tilde{X}^I_k(t)+\sqrt{\frac{N_k}{R}}x_0^I\, ,
\end{equation*}
where the rescaling of the zero-mode is a consequence of the field redefinition. The path integral over this mode then gives conservation of momentum in the target space
\begin{equation*}
    \int d^9x^I_0 \,e^{-ix_0^I(p_k^I-\Tilde{p}_k^I)}=(2\pi)^9\delta(\Vec{p}_k-\Tilde{\Vec{p}}_k) \, .
\end{equation*}
Now, the integral over $T$ can be thought of as an integral over a constant shift in time, and reads 
\begin{equation*}
   \int d\mu_{T}=  \int d\mu_{T}\,e^{-iT(p_k^- -\Tilde{p}_k^-)}= 2\pi \delta(p_k^- -\Tilde{p}_k^-) \, ,
\end{equation*}
where the first equality is trivially true since we are working on-shell, hence the delta is really a $\delta(0)$, which is what we expect for the energy conservation condition for on-shell eleven-dimensional momenta in the form (\ref{eq:dictionary}).
Hence, the remaining part of the Green's function reads 
\begin{equation}
\mathcal{A}(N_k, v_k, h;N_k, v_k, \Tilde{h})=\int d\mu_{t_k} \langle T[\mathcal{V}_k^{*}(0)\mathcal{V}_k(t_k)] \rangle \, .
\end{equation}
To compute this correlation function, we must consider all possible Wick contractions. There are three options, depending on how the operators in the prefactor are contracted.
\begin{itemize}
    \item \textbf{Option 1:} All prefactors are contracted together
    \begin{multline*}
    \mathcal{A}_1(N_k, v_k, h;N_k, v_k, \tilde{h})=2|\mathcal{N}_k|^2\int d\mu_{t_k} h_{IJ}\tilde{h}_{KL}
    \langle \dot{X}_k^I(0)\dot{X}_k^K(t_k) \rangle\langle \dot{X}_k^J(0)\dot{X}_k^L(t_k) \rangle \\
    \langle\cos{\Bigl(\beta_k\sqrt{\frac{2N_k}{R}} \int_{-\infty}^{+\infty} du \,\dot{X}_k(u)\cdot v_k\Bigr)}\cos{\Bigl(\beta_k\sqrt{\frac{2N_k}{R}} \int_{-\infty}^{+\infty} dw \,\dot{X}_k(w)\cdot v_k\Bigr)}\rangle \\
    \langle e^{-i\sqrt{\frac{2N_k}{R}}v_k \cdot X_k(0)}e^{i\sqrt{\frac{2N_k}{R}}v_k \cdot X_k(t_k)}\rangle
    \end{multline*}
    \item \textbf{Option 2:} One prefactor for each operator is contracted with the exponential
    \begin{multline*}
    \mathcal{A}_2(N_k, v_k, h;N_k, v_k, \tilde{h})=4|\mathcal{N}_k|^2\int d\mu_{t_k} h_{IJ}\tilde{h}_{KL}
    \langle \dot{X}_k^I(0)\dot{X}_k^K(t_k) \rangle \\
    \langle\cos{\Bigl(\beta_k\sqrt{\frac{2N_k}{R}} \int_{-\infty}^{+\infty} du \,\dot{X}_k(u)\cdot v_k\Bigr)}\cos{\Bigl(\beta_k\sqrt{\frac{2N_k}{R}} \int_{-\infty}^{+\infty} dw \,\dot{X}_k(w)\cdot v_k\Bigr)}\rangle \\
  \langle e^{-i\sqrt{\frac{2N_k}{R}}v_k \cdot X_k(0)}\dot{X}_k^L(t_k)\rangle \langle \dot{X}_k^J(0)e^{i\sqrt{\frac{2N_k}{R}}v_k \cdot X_k(t_k)}\rangle
    \end{multline*}
    \item \textbf{Option 3:} All prefactors are contracted with exponentials 
     \begin{multline*}
     \mathcal{A}_3(N_k, v_k, h;N_k, v_k, \tilde{h})=|\mathcal{N}_k|^2\int d\mu_{t_k} h_{IJ}\tilde{h}_{KL}\\
  \langle e^{-i\sqrt{\frac{2N_k}{R}}v_k \cdot X_k(0)}\dot{X}_k^K(t_k)\dot{X}_k^L(t_k)\rangle \langle \dot{X}_k^I(0)\dot{X}_k^J(0)e^{i\sqrt{\frac{2N_k}{R}}v_k \cdot X_k(t_k)}\rangle \\
  \langle\cos{\Bigl(\beta_k\sqrt{\frac{2N_k}{R}} \int_{-\infty}^{+\infty} du \,\dot{X}_k(u)\cdot v_k\Bigr)}\cos{\Bigl(\beta_k\sqrt{\frac{2N_k}{R}} \int_{-\infty}^{+\infty} dw \,\dot{X}_k(w)\cdot v_k\Bigr)}\rangle 
  \end{multline*}
\end{itemize}
After the prefactors are contracted, the final step is to contract the remaining exponentials, and the total amplitude is then the sum of the three contributions. The contraction between cosines reads
\[
\langle\cos{\Bigl(\beta_k\sqrt{\frac{2N_k}{R}} \int_{-\infty}^{+\infty} du \,\dot{X}_k(u)\cdot v_k\Bigr)}\cos{\Bigl(\beta_k\sqrt{\frac{2N_k}{R}} \int_{-\infty}^{+\infty} dw \,\dot{X}_k(w)\cdot v_k\Bigr)}\rangle =\cos{\Bigl(\frac{\beta^2_k}{\epsilon}\frac{N_k}{R}v_k^2\Bigr)}
\]
where we regularised a divergent delta function as the $\epsilon \rightarrow 0 $ limit of the right-hand side of (\ref{eq:twoaux}) when evaluated at the same time, so that $\delta(0)$ is replaced by $\frac{1}{2\epsilon}$. The first contribution is given by
\begin{equation}
\mathcal{A}_1(N_k, v_k, h;N_k, v_k, \tilde{h})=-\frac{|\mathcal{N}_k|^2}{\epsilon}h_{IJ}\tilde{h}^{IJ}\cos{\Bigl(\frac{\beta^2_k}{\epsilon}\frac{N_k}{R}v_k^2\Bigr)} \,.
\end{equation}
As mentioned above, the role of the cosine factor in the vertex operator must cancel the divergence that arises due to the one-dimensional, ultra-local nature of the theory. Hence, we choose
\begin{equation}
\label{eq:normcond}
\beta_k =\sqrt{\frac{R}{v^2_kN_k}\epsilon\,  \bigl(\frac{\pi}{2}+c_k\epsilon\bigr)}
\end{equation}
to cancel the divergence, where $c_k$ is a constant, which we will specify shortly. It follows that
\begin{equation}
    \label{eq:tern1}
    \mathcal{A}_1(N_k, v_k, h;N_k, v_k, \tilde{h})=|\mathcal{N}_k|^2c_k h_{IJ}\tilde{h}^{IJ} \, .
\end{equation}
Note that, in principle, we should have listed options where either the prefactor or the exponential in vertex-like operators is contracted with the cosine factor. However, options where the exponent is contracted with the prefactor are in the form
\[
\langle \int_{-\infty}^{+\infty} dudv\, \dot{X}^I(u) \dot{X}^J(v) X^K(t_k) X^L(t_k)\rangle=\int_{-\infty}^{+\infty} dudv \,\sgn(u-t_k)\sgn(v-t_k)=0 \, .
\]
Moreover, when velocity operators are contracted with the cosine, e.g. as an alternative to contracting prefactors with the exponent in option 3, the result is
\[
\langle \beta_k^2\frac{2N_k}{R}v_k^Iv_k^J\int_{-\infty}^{+\infty} dudv\, \dot{X}^I(u) \dot{X}^J(v) \dot{X}^K(t_k) \dot{X}^L(t_k)\rangle \propto \epsilon
\]
and vanishes in the $\epsilon \rightarrow 0$ limit. Hence, the only non-vanishing contraction of the cosine terms is among themselves, thus removing the divergence and leaving the rest of the amplitude unaffected. 

The remaining two options for the Wick's contractions read
\begin{equation}
\mathcal{A}_2(N_k, v_k, h;N_k, v_k, \tilde{h})=-i|\mathcal{N}_k|^2c_k \epsilon \frac{2N_k}{R}h_{IJ}v_k^J\tilde{h}^{IL}v_{k,L}
\label{eq:term2}
\end{equation}
and 
\begin{equation}
    \mathcal{A}_3(N_k, v_k, h;N_k, v_k, \tilde{h})=-\frac{i}{2}|\mathcal{N}_k|^2c_k \epsilon \frac{N_k}{R}\frac{h_{IJ}v_k^Iv_k^J\tilde{h}_{KL}v_k^Kv_k^L}{v_k^2} \, .
    \label{eq:term3}
\end{equation}
and vanish in the $\epsilon \rightarrow 0$ limit. It follows that only the first amplitude (\ref{eq:tern1}) contributes. Moreover, choosing
\begin{equation}
    \label{eq:normeps}
c_k=\frac{1}{|\mathcal{N}_k|^2}
\end{equation}
fixes the overall normalisation, and the final result for the amplitude reads
\begin{equation}
    \label{eq:gravamp}
    \mathcal{A}(N_k, v_k, h;N_k, v_k, \tilde{h})=\langle N_k,v_k,h|N_k, v_k, \tilde{h}\rangle =h_{IJ}\tilde{h}^{IJ}\,,
\end{equation}
which reproduces the overlap between graviton states.

We have thus shown that the two-point function of vertex-like operators reproduces the correct graviton amplitude in the target space. Indeed, (\ref{eq:gravamp}) implies that, if we attach an external line of a particle $j$ to an amplitude in the form (\ref{eq:Green}), we obtain, removing delta functions for momentum and energy conservation 
\begin{multline*}
    \int d\mu_{t_{ext}} \int d\mu_{\tilde{t}_j}\langle \mathcal{V}_j(t_{ext}) \tilde{\mathcal{V}}_{j, \alpha \beta}(\tilde{t}_j) \rangle\tilde{\mathcal{A}}^{\alpha \beta}(N_1, v^{I}_1, \theta_1;...;N_j, v^{I}_j;...; N_n, v^{I}_n, \theta_n)\\
    =\theta_{\alpha \beta}\tilde{\mathcal{A}}^{\alpha \beta}(N_1, v^{I}_1, \theta_1;...;N_j, v^{I}_j;...; N_n, v^{I}_n, \theta_n) \, ,
\end{multline*}
where $\Tilde{\mathcal{A}}^{\alpha \beta}(N_1, v^{I}_1, \theta_1;...;N_j, v^{I}_j;...; N_n, v^{I}_n, \theta_n)$ is the tensor obtained by considering a generic scattering amplitude between vertex-like operators, with the external graviton polarisation removed. Similarly, the notation $\Tilde{\mathcal{V}}_{j, \alpha \beta}(\Tilde{t}_j)$ stands for a graviton operator which is not contracted to an external polarisation, such that the two-point function 
\begin{equation*}
    \int d\mu_{t_{ext}} \int d\mu_{\Tilde{t}_j}\langle \mathcal{V}_j(t_{ext}) \Tilde{\mathcal{V}}_{j, \alpha \beta}(\Tilde{t}_j) \rangle
\end{equation*} 
simply reproduces the external graviton polarisation $\theta_{\alpha \beta}$.

We are now ready to use this construction of vertex-like operators to show the soft factorisation property at leading and subleading orders.
\section{The leading soft term}
\label{sec:leading}
In this section, we will show how to reproduce the leading term of the soft factorisation (\ref{eq:softtheorem}) from the formalism outlined in section \ref{sec:setup}.  We will carry out the computation in detail, thus providing a proof for the soft theorem for $D_0$ bound states in BFSS at leading order in the large $N$ expansion. 

 We consider the amplitude (\ref{eq:ampint}), where we make vertex operators explicit. As explained in section \ref{sec:vertop}, vertex operators (\ref{eq:vertnorm}) have a cosine term which allows to remove a divergence in the graviton two-point function. However, we saw that they play no role in Wick's contractions, since they give rise to vanishing terms when $\epsilon \rightarrow 0$. Hence, we will just set them equal to $1$ in the following, for ease of notation. The amplitude (\ref{eq:ampint}) reads
\begin{multline}
    \mathcal{A}(N_s, v^{I}_s, h_s;N_1, v^{I}_1, \theta_1;...; N_n, v^{I}_n, \theta_n) \\
    =\mathcal{N}_s^*\mathcal{N}_j^*\int d\mu_{t_s}\prod_{i=1}^n \int d\mu_{t_i} \langle \Omega| h_{IJ}T[\dot{X}^{I}_s(t_s)\dot{X}_s^{J}(t_s) e^{-i \sqrt{\frac{2N_s}{R}}v_s \cdot X_s(t_s)} \\
   \times \sum_{N=1}^{\infty} \, \left[\frac{i^N}{N!}\Bigl( \int du\,\mathcal{L}_{s j} (u)\Bigr)^N\right] A_{I_1 I_2} \dot{X}_j^{I_1}(t_j) \dot{X}_j^{I_2}(t_j) e^{-i \sqrt{\frac{2N_j}{R}}v_j \cdot X_j(t_j)}\prod_{k \neq j}\mathcal{V}_k(t_k)]|\Omega\rangle\, ,
   \label{eq:Osh}
\end{multline}
where, for simplicity, we assumed the hard particle to be also a graviton with polarisation $A_{I_1 I_2}$, and it is understood that in $\prod_{k \neq j}\mathcal{V}_k(t_k)$ operators corresponding to outgoing particles are complex conjugated. As explained in section \ref{sec:vertop}, the integral over zero-modes of fields yields a delta function for total momentum conservation. 
We need to evaluate the various options for Wick's contractions involving the soft vertex operator, the $j$-th hard vertex operators and interaction terms in (\ref{eq:Osh}). We can make a distinction between contractions among interaction terms in $\prod_i \mathcal{L}_{sj}(u_i)$, which we will name internal contractions; and contractions between operators in $\prod_i \mathcal{L}_{sj}(u_i)$ and terms in the external vertex operators, which we will name external contractions. It will be useful to first work out the result of internal contractions. 

First we rewrite (\ref{eq:Osh}) as 
\begin{multline}
   \mathcal{A}(N_s, v^{I}_s, h_s;N_1, v^{I}_1, \theta_1;...; N_n, v^{I}_n, \theta_n)\\
   =\mathcal{N}_s^*\mathcal{N}_j^*\int d\mu_{t_s}\prod_{i=1}^n \int d\mu_{t_i}\langle \Omega| h_{IJ}T[\dot{X}^{I}_s(t_s)\dot{X}_s^{J}(t_s) e^{-i \sqrt{\frac{2N_s}{R}}v_s \cdot X_s(t_s)} \\
   \times \sum_{N=1}^{\infty} \, \left[\frac{i^N}{N!} \sum_{k=1}^N \,\sum_{n_1+...+n_k=N}\,\frac{N!}{k!n_1! \, ... \,  n_k!}\prod_{i=1}^k\Bigl(\int du\,\mathcal{L}_{s j} (u)\Bigr)^{n_i}\right] \\
   \times A_{I_1 I_2} \dot{X}_j^{I_1}(t_j) \dot{X}_j^{I_2}(t_j) e^{-i \sqrt{\frac{2N_j}{R}}v_j \cdot X_j(t_j)}\prod_{k \neq j}\mathcal{V}_k(t_k)]|\Omega \rangle\, .
\end{multline}
Next, we rearrange the sums as 
\begin{equation}
    \sum_{N=1}^{\infty}\,\sum_{k=1}^N \,\sum_{n_1+...+n_k=N}=\sum_{k=1}^{\infty} \, \sum_{N=k}^{\infty} \, \sum_{n_1+...+n_k=N}=\sum_{k=1}^{\infty} \,\prod_{i=1}^k \sum_{n_i=1}^{\infty} \, ,
\end{equation}
thus obtaining
\begin{multline}
   \mathcal{A}(N_s, v^{I}_s, h_s;N_1, v^{I}_1, \theta_1;...; N_n, v^{I}_n, \theta_n)\\
   =\mathcal{N}_s^*\mathcal{N}_j^*\int d\mu_{t_s}\prod_{i=1}^n \int d\mu_{t_i}\langle \Omega| h_{IJ}T[\dot{X}^{I}_s(t_s)\dot{X}_s^{J}(t_s) e^{-i \sqrt{\frac{2N_s}{R}}v_s \cdot X_s(t_s)} \\
   \times \left[\sum_{k=1}^{\infty}\frac{1}{k!} \,\prod_{i=1}^k \sum_{n_i=1}^{\infty}\,\frac{1}{n_i!}\Bigl(i\int du\,\mathcal{L}_{s j} (u)\Bigr)^{n_i}\right]\\
   \times A_{I_1 I_2} \dot{X}_j^{I_1}(t_j)\dot{X}_j^{I_2}(t_j) e^{-i \sqrt{\frac{2N_j}{R}}v_j \cdot X_j(t_j)}\prod_{k \neq j}\mathcal{V}_k(t_k)]| \Omega\rangle\, .
   \label{eq:G2}
\end{multline}
Let us define a chain of length $n$ as 
\begin{equation}
    C_{n}=\Bigl(-\frac{i}{2}\Bigr)^{n}  \langle\prod_{i=1}^{n} \int du_i \, \bigl(\lambda(u_i)\sigma(u_i)\bigr) \Bigl(\sqrt{\frac{R}{N_s}}\Dot{X}_s^I(u_i)-\sqrt{\frac{R}{N_j}}\Dot{X}_j^I(u_i)\Bigr)^2 \rangle_{\omega}
\end{equation}
where the angle brackets $\langle \rangle_w$ mean that all $\Dot{X}_j$ operators except two are contracted among each other, in such a way that the contraction only involves operators evaluated at different times. Soft operators do not take part in internal contractions, since we are only considering the case in which the soft particle appears in external lines (soft emission or absorption). The two operators that remain uncontracted are called the endpoints of the chain and will take part in external contractions. 
For example, consider the $n=2$ case:
\begin{multline*}
    \int du_1 du_2\frac{(-i) \lambda(u_1)}{2} T[\sigma(u_1) \Bigl(\sqrt{\frac{R}{N_s}}\Dot{X}_s^I(u_1)-\sqrt{\frac{R}{N_j}}\Dot{X}_j^I(u_1)\Bigr)\Bigl(\underline{\sqrt{\frac{R}{N_s}}\Dot{X}_s^I(u_1)-\sqrt{\frac{R}{N_j}}\Dot{X}_j^I(u_1)}\Bigr)\\
    \times \frac{(-i) \lambda(u_2)}{2} \sigma(u_2) \Bigl(\underline{\sqrt{\frac{R}{N_s}}\Dot{X}_s^J(u_2)-\sqrt{\frac{R}{N_j}}\Dot{X}_j^J(u_2)}\Bigr)\Bigl(\sqrt{\frac{R}{N_s}}\Dot{X}_s^J(u_2)-\sqrt{\frac{R}{N_j}}\Dot{X}_j^J(u_2)\Bigr) ]\times 4\, ,
\end{multline*}
where the underlined terms are contracted. The factor of $4$ accounts for the equivalent possibilities. This evaluates to 
\begin{equation*}
  -i \int du_1\bigl[ (\lambda(u_1) \sigma(u_1) \bigr]^2\frac{R}{N_j}\Bigl(\sqrt{\frac{R}{N_s}}\dot{X}_s^I(u_1)-\sqrt{\frac{R}{N_j}}\dot{X}_j^I(u_1)\Bigr)^2\,.
\end{equation*}
In general, performing the contraction of the $\Dot{X}_j$ operators yields
\begin{equation}
    C_{n}=-in!\frac{N_j}{R}\Bigl(-\frac{i}{2}\Bigr)^{n}\Bigl(\frac{2iR}{N_j}\Bigr)^n \int du \, \lambda^n(u)\sigma^n(u)\Bigl(\sqrt{\frac{R}{N_s}}\dot{X}_s^I(u)-\sqrt{\frac{R}{N_j}}\dot{X}_j^I(u)\Bigr)^2 \, .
    \label{eq:chaininteraction}
\end{equation}
Now define 
\begin{equation}
    F[\sigma(u)]=\sum_{n=1}^{\infty}\Bigl(\frac{\sigma(u) \lambda(u)R}{N_j}\Bigr)^n=\frac{\sigma(u) \lambda(u) \frac{R}{N_j}}{1- \sigma(u) \lambda(u) \frac{R}{N_j}} \, ,
    \label{eq:Fop}
\end{equation}
so that 
\begin{equation}
    \sum_{n=1}^{\infty}\,\frac{1}{n!}C_{n}=-i \frac{N_j}{R} \int du \, F[\sigma(u)] \Bigl(\sqrt{\frac{R}{N_s}}\dot{X}_s^I(u)-\sqrt{\frac{R}{N_j}}\dot{X}_j^I(u)\Bigr)^2 \, ,
\end{equation}
and for convenience, introduce 
\begin{equation}
    O_{res}:=\left[\sum_{k=1}^{\infty}\frac{1}{k!} \,\prod_{i=1}^k \sum_{n_i=1}^{\infty}\,\frac{1}{n_i!}C_{n_i}\right] \, .
    \label{eq:Ores}
\end{equation}
Then (\ref{eq:G2}) becomes
\begin{multline}
   \mathcal{A}(N_s, v^{I}_s, h_s;N_1, v^{I}_1, \theta_1;...; N_n, v^{I}_n, \theta_n)\\
   =\mathcal{N}_s^*\mathcal{N}_j^*\int d\mu_{t_s}\prod_{i=1}^n \int d\mu_{t_i}  \langle \Omega| h_{IJ}T[\dot{X}^{I}_s(t_s)\dot{X}_s^{J}(t_s) e^{-i \sqrt{\frac{2N_s}{R}}v_s \cdot X_s(t_s)} \\
   \times O_{res} A_{I_1 I_2} \dot{X}_j^{I_1}(t_j)\dot{X}_j^{I_2}(t_j) e^{-i \sqrt{\frac{2N_j}{R}}v_j \cdot X_j(t_j)}\prod_{k \neq j}\mathcal{V}_k(t_k)]|\Omega \rangle\, .
   \label{eq:amp2}
\end{multline}
The next step is to work out external contractions involving the endpoints of the chains. 
In order to provide a non-zero contribution to leading (or first sub-leading) order in $N$,  external contractions must satisfy:
\begin{enumerate}[label=\Alph*)]
    \item Every remaining velocity operator in the chains, and all auxiliary fields $\sigma$, must be contracted, because otherwise the term will be killed by normal ordering;
    \item The soft operators $\dot X_s(t_s)$  
    must be contracted out, each with a different chain.
    This is because, by (\ref{eq:twovel}), contracting the two soft operators in (\ref{eq:amp2}) with the \emph{same} chain gives a term proportional to $h_{IJ} \delta^{IJ}$ which vanishes because the graviton polarisation is traceless; it follows that the $k=1$ term in \eqref{eq:Ores} does not contribute. For the same reason the two hard velocity operators, $\dot X_j(t_j)$, cannot be contracted with the same chain;
    \item At least one end of each chain must be contracted with an exponential operator. This is because terms in which $\dot X_s(t_s)$ is contracted with one end of a chain and $\dot X_j(t_j)$ with the other end lead to a factor $\delta(t_s-t_j)$ which kills the integral over $t_s$. 
    As we will see in the explicit computation below, this implies that the amplitude is proportional to $\lambda(t_j)\sim \frac{1}{N^2}$, which is lower order with respect to subleading in $N$. Hence, we neglect these terms. 
\end{enumerate}
We can write down a Feynman rule for the external edges of a diagram corresponding to contractions with an external velocity: 
\begin{multline}
    \langle :\dot{X}^I_l(t)e^{-i\Bigl( \sqrt{\frac{2N_s}{R}}v_s \cdot X_s(t_s)+\sqrt{\frac{2N_j}{R}}v_j \cdot X_j(t_j)\Bigr)}:\sum_{n=1}^{\infty}\,\frac{1}{n!}C_{n}\rangle\\
     =-\frac{N_j}{\sqrt{2}R}\sqrt{\frac{R}{N_l}}v_{sj}^IF[\sigma(t)]e^{-i\Bigl( \sqrt{\frac{2N_s}{R}}v_s \cdot X_s(t_s)+\sqrt{\frac{2N_j}{R}}v_j \cdot X_j(t_j)\Bigr)}\sum_{n=1}^{\infty}\,\frac{1}{n!}C_{n}
     \label{eq:Feynrule}
\end{multline}
where $l=s,j$ depending on whether the external velocity is soft or hard, and we defined $v^I_{sj}=(v^I_s-v^I_j)$. 
In particular, this implies an $F[\sigma]$ power counting for external contractions, and we will show in the following how this translates into a power counting in $N$. 

There are three possibilities for the external contractions, and we specify for each of them the power of $F[\sigma]$ resulting from contractions with external velocities:
\begin{enumerate}
    \item No operator from the hard vertex operator is contracted, 
    \begin{multline}
      \mathcal{A}^{(1)}(N_s, v^{I}_s, h_s;N_1, v^{I}_1, \theta_1;...; N_n, v^{I}_n, \theta_n)\\
   =\mathcal{N}_s^*\mathcal{N}_j^*\int d\mu_{t_s}\prod_{i=1}^n \int d\mu_{t_i} h_{IJ}
      \langle \Omega| T[\underline{\dot{X}_s^{I}(t_s)\dot{X}_s^{J}(t_s) e^{-i\Bigl( \sqrt{\frac{2N_s}{R}}v_s \cdot X_s(t_s)+\sqrt{\frac{2N_j}{R}}v_j \cdot X_j(t_j)\Bigr)} } 
      \\
      \times \underline{O_{res} }A_{I_1 I_2}\dot{X}_j^{I_1}(t_j) \dot{X}_j^{I_2}(t_j)\prod_{k \neq j}\mathcal{V}_k(t_k)] |\Omega\rangle \, ,
    \label{eq:nohardcontr}
    \end{multline}
    where the underline indicates Wick contractions. The remaining string of vertex operators $\mathcal{V}^{*}_1(t_1)...\mathcal{V}_n(t_n)$ excludes the $j$-th one, which has been made explicit. The power counting yields $F^2[\sigma(t_s)]$;
    \item One of the operators from the hard vertex operator is contracted,
    \begin{multline}
      \mathcal{A}^{(2)}(N_s, v^{I}_s, h_s;N_1, v^{I}_1, \theta_1;...; N_n, v^{I}_n, \theta_n)\\
   =\mathcal{N}_s^*\mathcal{N}_j^*\int d\mu_{t_s}\prod_{i=1}^n \int d\mu_{t_i} 
      h_{IJ}\langle \Omega| T[\underline{\dot{X}_s^{I}(t_s)\dot{X}_s^{J}(t_s) e^{-i\Bigl( \sqrt{\frac{2N_s}{R}}v_s \cdot X_s(t_s)+\sqrt{\frac{2N_j}{R}}v_j \cdot X_j(t_j)\Bigr)} }\\
    \times \underline{O_{res} }A_{I_1 I_2} \underline{\dot{X}_j^{I_1}(t_j)} \dot{X}_j^{I_2}(t_j) \prod_{k \neq j}\mathcal{V}_k(t_k)]|\Omega\rangle \, .
    \label{eq:1hardcontr}
    \end{multline}
    The power counting yields $F^2[\sigma(t_s)]F[\sigma(t_j)]$;
    \item Both operators from the hard vertex operator are contracted,
    \begin{multline}
     \mathcal{A}^{(3)}(N_s, v^{I}_s, h_s;N_1, v^{I}_1, \theta_1;...; N_n, v^{I}_n, \theta_n)\\
   =\mathcal{N}_s^*\mathcal{N}_j^*\int d\mu_{t_s}\prod_{i=1}^n \int d\mu_{t_i}
     h_{IJ}\langle\Omega| T[\underline{\dot{X}_s^{I}(t_s)\dot{X}_s^{J}(t_s) e^{-i\Bigl( \sqrt{\frac{2N_s}{R}}v_s \cdot X_s(t_s)+\sqrt{\frac{2N_j}{R}}v_j \cdot X_j(t_j)\Bigr)} }\\
  \times \underline{O_{res}}A_{I_1 I_2} \underline{\dot{X}_j^{I_1}(t_j)\dot{X}_j^{I_2}(t_j)}\prod_{k \neq j}\mathcal{V}_k(t_k)]|\Omega\rangle \, .
     \label{eq:2hardcontr}
    \end{multline}
     The power counting yields $F^2[\sigma(t_s)]F^2[\sigma(t_j)]$.
\end{enumerate}
Note that we only highlighted the $F[\sigma]$ power counting stemming from contractions with external operators. There are $F[\sigma]$ factors coming from contractions of chains in $O_{res}$ with the exponential operator. Those are the same for all three contributions and will be worked out explicitly in the following. We will compute the case (\ref{eq:nohardcontr}) explicitly; the others are similar, and we will point out how the $N$ power counting arises from the corresponding $F[\sigma]$ power counting as we go through the steps of the computation. 

Firstly, the $k=0$ term in (\ref{eq:Ores}) does not contribute because of rule C and the $k=1$ term vanishes due to rule B. This implies that non-zero contributions are those for which the sum over $k$ in (\ref{eq:Ores}) starts from two, namely
\begin{equation}
    O_{res}=\left[\sum_{k=2}^{\infty}\frac{1}{k!} \,\prod_{i=1}^k \sum_{n_i=1}^{\infty}\,\frac{1}{n_i!}C_{n_i}\right] \, .
\end{equation}
We select two chains to be contracted with the soft vertex operator. There are $\frac{k(k-1)}{2}$ equivalent options for this choice. We then contract them with external soft operators. After this operation, we obtain
\begin{multline}
     \mathcal{A}(N_s, v^{I}_s, h_s;N_1, v^{I}_1, \theta_1;...; N_n, v^{I}_n, \theta_n)\\
   =\mathcal{N}_s^*\mathcal{N}_j^*\int d\mu_{t_s}\prod_{i=1}^n \int d\mu_{t_i}
    \langle \Omega |h_{IJ}T[v^{I}_{sj} v^{J}_{sj} e^{-i\sqrt{\frac{2N_s}{R}}v_s \cdot X_s(t_s)}\Bigl(\frac{N_j}{R}\Bigr)^2\Bigl(\frac{R}{N_s}\Bigr)\\ \times F^2[\sigma(t_s)]\left[\sum_{k=0}^{\infty}\frac{1}{k!} \,\prod_{i=1}^k \sum_{n_i=1}^{\infty}\,\frac{1}{n_i!}C_{n_i}\right]A_{I_1 I_2} \dot{X}_{j}^{I_1}(t_j) \dot{X}_j^{I_2}(t_j) e^{-i \sqrt{\frac{2N_j}{R}}v_j \cdot X_j(t_j)}\prod_{k \neq j}\mathcal{V}_k(t_k)]|\Omega \rangle\, .
   \label{eq:G4}
\end{multline}
Note that in the intermediate steps of the previous computation, factors of $\delta(u_{1,2}-t_s)$ appear due to (\ref{eq:twovel}). Then equation (\ref{eq:dysonseries}) tells us that, upon integration over $u_{1,2}$, this is non-vanishing only for $t_j \geq t_s$ (it is the other way around for two incoming particles, whereas for one \textit{incoming} and one \textit{outgoing} particle indeed the \textit{outgoing} one comes later). Note that the Feynman rule (\ref{eq:Feynrule}) has been applied in order to obtain contractions of chains with external operators, and the factor $F^2[\sigma(t_s)]$ has been obtained, as anticipated. The analogous application of the Feynman rule (\ref{eq:Feynrule}) to the terms (\ref{eq:1hardcontr}) and (\ref{eq:2hardcontr}) yields the respective $F[\sigma]$ powers described above.

The next step consists of working out the $\sigma$ contractions. A contraction of $\sigma$ fields, in the $\epsilon \rightarrow 0$ limit, reads 
\begin{equation}
    \langle \sigma(u_1) \sigma(u_2) \rangle= \frac{i}{\epsilon \sqrt{\lambda(u_1)\lambda(u_2)}} e^{-\frac{|u_1-u_2|}{\epsilon}} \rightarrow \frac{2i}{\lambda(u_1)} \delta(u_1-u_2) \, .
    \label{eq:auxdelta}
\end{equation}
 in particular, propagator between two $\sigma$'s is proportional to $\frac{1}{\lambda}$. Moreover, we recall the definition (\ref{eq:Fop}) of $F[\sigma(t)]$ operators, which implies that every $\sigma$ comes in the combination $\sigma(t)\lambda$. This, in turn, implies that for any pair of contracted $\sigma$'s, we get a factor of $\lambda \sim O(\frac{1}{N^2})$. For example, we can compute 
 \begin{equation}
     \label{eq:F}
     \langle F[\sigma(t_s)]\rangle=\frac{i}{\epsilon}\Bigl(\frac{R}{N_j}\Bigr)^2\lambda(t_s)\Bigl(1+O\bigl(\lambda(t_s)\bigr)\Bigr)=\frac{i}{\epsilon}\Bigl(\frac{R}{N_j}\Bigr)^2 \lambda(t_s)\Bigl(1+O(\frac{1}{N^2})\Bigr)
 \end{equation}
 and 
\begin{equation}
    \langle F[\sigma(t_s)]F[\sigma(t_s)]\rangle=\frac{i}{\epsilon}\Bigl(\frac{R}{N_j}\Bigr)^2 \lambda(t_s)\Bigl(1+O(\frac{1}{N^2})\Bigr) \, .
    \label{eq:F2}
\end{equation}
We can work out the analogous contraction for $F[\sigma]$ powers in the terms (\ref{eq:1hardcontr}) and (\ref{eq:2hardcontr}). These read
\begin{equation}
    \langle F[\sigma(t_s)]F[\sigma(t_s)]F[\sigma(t_j)]\rangle=-2\Bigl(\frac{R}{N_j}\Bigr)^4 \frac{\lambda(t_s)}{\epsilon}\frac{\lambda(t_j)}{\epsilon}\Bigl(1+O(\frac{1}{N^2})\Bigr)
    \label{eq:F3}
\end{equation}
and
\begin{equation}
    \langle F[\sigma(t_s)]F[\sigma(t_s)]F[\sigma(t_j)] F[\sigma(t_j)]\rangle=-2\Bigl(\frac{R}{N_j}\Bigr)^4 \frac{\lambda(t_s)}{\epsilon}\frac{\lambda(t_j)}{\epsilon}\Bigl(1+O(\frac{1}{N^2})\Bigr) \,.
    \label{eq:F4}
\end{equation}
In particular, it is clear from (\ref{eq:F3}) and (\ref{eq:F4}) that those contributions are $O(\frac{1}{N^2})$ with respect to (\ref{eq:F2}). This implies that the external contractions (\ref{eq:1hardcontr}) and (\ref{eq:2hardcontr}) are suppressed with respect to (\ref{eq:nohardcontr}) and do not contribute to the soft theorem. Note that this is true even though the amplitudes they multiply are different. Those amplitudes involve only hard particles and are computed in the full theory, which is strongly coupled in the large $N$ limit. Hence, correlators differing by $\dot{X}_j$ factors do not change the polynomial scaling with $N$ of the prefactors. They are not affected by the soft limit and, as a consequence, do not change the above scalings.

We plug (\ref{eq:F2}) back into (\ref{eq:G4}), keeping only the leading order in the $O(\frac{1}{N^2})$ expansion, obtaining
\begin{multline}
   \mathcal{A}(N_s, v^{I}_s, h_s;N_1, v^{I}_1, \theta_1;...; N_n, v^{I}_n, \theta_n)\\
   =i\mathcal{N}_s^*\mathcal{N}_j^* \int d\mu_{t_s}\prod_{i=1}^n \int d\mu_{t_i}
    \langle \Omega |h_{IJ}T[v^{I}_{sj} v^{J}_{sj}\, e^{-i \sqrt{\frac{2N_s}{R}}v_s \cdot X_s(t_s)}\,\frac{\lambda(t_s)}{\epsilon}\Bigl(\frac{R}{N_s}\Bigr) \\
   \times \left[\sum_{k=0}^{\infty}\frac{1}{k!} \,\prod_{i=1}^k \sum_{n_i=1}^{\infty}\,\frac{1}{n_i!}C_{n_i}\right] A_{I_1 I_2} \dot{X}_{j}^{I_1}(t_j) X_j^{I_2}(t_j) e^{-i \sqrt{\frac{2N_j}{R}}v_j \cdot X_j(t_j)}\prod_{k \neq j}\mathcal{V}_k(t_k)]|\Omega \rangle \, .
   \label{eq:Amp1}
\end{multline}  
  The next step is to contract the remaining chains with the exponential operator. When doing this, the velocity operator factor can be contracted straightforwardly, giving 
\begin{multline}
  \mathcal{A}(N_s, v^{I}_s, h_s;N_1, v^{I}_1, \theta_1;...; N_n, v^{I}_n, \theta_n)\\
   =i\mathcal{N}_s^*\mathcal{N}_j^* \int d\mu_{t_s}\prod_{i=1}^n \int d\mu_{t_i}\langle \Omega |h_{IJ}T[v^{I}_{sj} v^{J}_{sj} e^{-i \sqrt{\frac{2N_s}{R}}v_s \cdot X_s(t_s)}\frac{\lambda(t_s)}{\epsilon}\Bigl(\frac{R}{N_s}\Bigr) \\
   \times \left[\sum_{k=0}^{\infty}\frac{1}{k!} \,\prod_{i=1}^k \Bigl(-i\frac{N_j}{2R}v^2_{sj}\Bigr)\int_{-\infty}^{t_s} du_i \,F[(\sigma_{u_i})]\right] A_{I_1 I_2} \dot{X}_{j}^{I_1}(t_j) \dot{X}_j^{I_2}(t_j) e^{-i \sqrt{\frac{2N_j}{R}}v_j \cdot X_j(t_j)}\prod_{k \neq j}\mathcal{V}_k(t_k)]|\Omega \rangle \, .
\end{multline}
Next, the remaining $\sigma$'s must be contracted among themselves. In order to do this, we use (\ref{eq:F}) and only keep the dominant contribution in the $O(\frac{1}{N^2})$ expansion. The expectation value we need to compute takes the form $\langle e^{\kappa\int_{-\infty}^{t_s} du_i \,F[(\sigma_{u_i})]}\rangle$. We want to show that
\begin{equation} 
\label{eq:expevaluation}
\langle e^{\kappa\int_{-\infty}^{t_s} du_i \,F[(\sigma_{u_i})]}\rangle=e^{\kappa\int_{-\infty}^{t_s} du_i \,\langle F[(\sigma_{u_i})]\rangle}(1+O(\epsilon)) \, .
\end{equation}
In order to do so, we consider a term at order $n$ in the expansion of $e^{\kappa\int_{-\infty}^{t_s} du_i \,\langle F[(\sigma_{u_i})]\rangle}$
\begin{multline*}
    e^{\kappa\int_{-\infty}^{t_s} du_i \,\langle F[(\sigma_{u_i})]\rangle} \supset \frac{\kappa^n}{n!}\int du_1...du_n \langle F[\sigma(u_1)]\rangle... \langle F[\sigma(u_n)] \rangle \\
    = \frac{\kappa^n}{n!}\int du_1...du_n \sum_{\substack{m_1,...,m_n=0 \\ m=m_1+...m_n}}^{\infty}c_{m_1}...c_{m_n} \Bigl(\frac{R}{N_j}\Bigr)^{2(n+m)}\Bigl(\frac{\lambda(u_1)}{\epsilon}\Bigr)^{1+m_1}... \Bigl(\frac{\lambda(u_n)}{\epsilon}\Bigr)^{1+m_n}
\end{multline*}
where we expanded $F$ contractions to all orders with combinatorial coefficients $c_{m_i}$ (in the rest of the computation below, we will be interested in the leading order in $\lambda$). Now we consider a term at order $n+1$ in the expansion of $\langle e^{\kappa\int_{-\infty}^{t_s} du_i \,F[(\sigma_{u_i})]}\rangle-e^{\kappa\int_{-\infty}^{t_s} du_i \,\langle F[(\sigma_{u_i})]\rangle}$
\begin{multline*}
    \langle e^{\kappa\int_{-\infty}^{t_s} du_i \,F[(\sigma_{u_i})]}\rangle-e^{\kappa\int_{-\infty}^{t_s} du_i \,\langle F[(\sigma_{u_i})]\rangle} \\
    \supset \frac{\kappa^{n+1}}{(n+1)!} \int du_1...du_{n+1}\, (n+1)\langle F[\sigma(u_1)] \rangle... \langle F[\sigma(u_n)] F[\sigma(u_{n+1})] \rangle
\end{multline*}
where we evaluate a contraction between two $F$ functionals evaluated at different times. The combinatorial factor is $n+1$ because time integrations are ordered, and we can only contract adjacent functionals together. The contraction evaluates to 
\[
\langle F[\sigma(u_n)] F[\sigma(u_{n+1})] \rangle=2\Bigl(\frac{R}{N_j}\Bigr)^2\delta(u_n-u_{n+1})\lambda(u_n) \sum_{m_n=0}^\infty\tilde{c}_{m_n}\Bigl(\frac{R}{N_j}\Bigr)^{2m_n}\Bigl(\frac{\lambda(u_n)}{\epsilon}\Bigr)^{m_n}
\]
so one time integration drops out with the delta function. We can group the two contributions together, leading to
\begin{multline*}
\langle e^{\kappa\int_{-\infty}^{t_s} du_i \,F[(\sigma_{u_i})]}\rangle \\
\supset \frac{\kappa^n}{n!}\int du_1...du_n \sum_{m_1,...,m_n=0}^{\infty}c_{m_1}...c_{m_{n-1}}(c_{m_n}-2\kappa \epsilon \tilde{c}_{m_n}) \Bigl(\frac{R}{N_j}\Bigr)^{2(n+m)}\Bigl(\frac{\lambda(u_1)}{\epsilon}\Bigr)^{1+m_1}... \Bigl(\frac{\lambda(u_n)}{\epsilon}\Bigr)^{1+m_n}
\end{multline*}
and we see that the term coming from contractions between different $F$ is $O(\epsilon)$ compared to the one coming from $e^{\kappa\int_{-\infty}^{t_s} du_i \,\langle F[(\sigma_{u_i})]\rangle}$. Indeed, it is a general fact that, whenever we consider a contraction between two different $F$, we generate a delta function which eliminates a time integral, and we pay the price of a power of $\epsilon$. The argument is generic in $n$ and can be repeated to all orders in the expansion of the exponential, proving the statement (\ref{eq:expevaluation}).

As a result, the amplitude reads
\begin{multline*}
  \mathcal{A}(N_s, v^{I}_s, h_s;N_1, v^{I}_1, \theta_1;...; N_n, v^{I}_n, \theta_n)\\
   =i\mathcal{N}_s^*\mathcal{N}_j^* \int d\mu_{t_s}\prod_{i=1}^n \int d\mu_{t_i}\langle \Omega |h_{IJ}T[v^{I}_{sj} v^{J}_{sj} e^{-i \sqrt{\frac{2N_s}{R}}v_s \cdot X_s(t_s)}\,\frac{\lambda(t_s)}{\epsilon} \Bigl(\frac{R}{N_s}\Bigr) \\
   \times  e^{\frac{R}{2N_j}v^2_{sj}\int_{-\infty}^{t_s} du_i \, \frac{\lambda(u_i)}{\epsilon}} A_{I_1 I_2}\dot{X}_{j}^{I_1}(t_j) \dot{X}_j^{I_2}(t_j) e^{-i\sqrt{\frac{2N_j}{R}}v_j \cdot X_j(t_j)}\prod_{k \neq j}\mathcal{V}_k(t_k)]|\Omega \rangle \, .
\end{multline*}
The expression for the amplitude can be further simplified by noting that the soft exponent is not involved in any additional contraction and is, therefore, set to 1 by normal ordering, yielding 
\begin{multline}
  \mathcal{A}(N_s, v^{I}_s, h_s;N_1, v^{I}_1, \theta_1;...; N_n, v^{I}_n, \theta_n)
  \\
  =i\mathcal{N}_s^*\mathcal{N}_j^*\int d\mu_{t_s}\prod_{i=1}^n \int d\mu_{t_i}\langle \Omega |h_{IJ}T[v^{I}_{sj} v^{J}_{sj} \frac{\lambda(t_s)}{\epsilon} \Bigl(\frac{R}{N_s}\Bigr)\\
   \times  e^{\frac{R}{2N_j}v^2_{sj}\int_{-\infty}^{t_s} du_i \, \frac{\lambda(u_i)}{\epsilon}} A_{I_1 I_2} \dot{X}_{j}^{I_1}(t_j) \dot{X}_j^{I_2}(t_j) e^{-i \sqrt{\frac{2N_j}{R}}v_j \cdot X_j(t_j)}\prod_{k \neq j}\mathcal{V}_k(t_k)]|\Omega \rangle \, .
   \label{eq:A5}
\end{multline}
We can now carry out the integration over $t_s$. The integral takes the form
\begin{equation}
    \int d\mu_{t_s} \frac{\lambda(t_s)}{\epsilon} \, e^{\frac{R}{2N_j}v^2_{sj}\int_{-\infty}^{t_s} du_i \, \frac{\lambda(u_i)}{\epsilon}}\\
    =\lim_{\alpha \rightarrow 0}\int_{-\infty}^{ t_j} dt_s\,\frac{\lambda(t_s)}{\epsilon}\, e^{\frac{R}{2N_j}v^2_{sj}\int_{-\infty}^{t_s} du_i \, \frac{\lambda(u_i)}{\epsilon}} e^{-\alpha |t_s|}\, .
\end{equation}
We evaluate the integral under the change of variable 
\begin{equation*}
    k(t_s)=\int_{-\infty}^{t_s} du_i \, \frac{\lambda(u_i)}{\epsilon}-\frac{N_j}{R}\frac{2\alpha}{v^2_{sj}} |t_s|
\end{equation*} 
and neglecting a term that goes to zero with $\alpha$. We obtain 
\begin{equation}
    \int d\mu_{t_s}\frac{\lambda(t_s)}{\epsilon} \, e^{\frac{R}{2N_j}v^2_{sj}\int_{-\infty}^{t_s} du_i \, \frac{\lambda(u_i)}{\epsilon}}= \int_{-\infty}^{k(t_j)} dk\, e^{\frac{R}{2N_j}v^2_{sj} \, k}=\frac{N_j}{R}\frac{2}{v^2_{sj}} \Bigl(1+O(\frac{1}{N^2})\Bigr) \, .
\end{equation}
By plugging back into equation (\ref{eq:A5}) we obtain
\begin{multline}
     \mathcal{A}(N_s, v^{I}_s, h_s;N_1, v^{I}_1, \theta_1;...; N_n, v^{I}_n, \theta_n)
  =2i\mathcal{N}_s^*\mathcal{N}_j^*\Bigl(\frac{N_j}{N_s}\Bigr)h_{IJ}\frac{v^{I}_{sj} v^{J}_{sj}}{v^2_{sj}} \\\
  \prod_{i=1}^n \int d\mu_{t_i}\langle \Omega |T[A_{I_1 I_2} \dot{X}_{j}^{I_1}(t_j) \dot{X}_j^{I_2}(t_j) e^{-i\sqrt{\frac{2N_j}{R}}v_j \cdot X_j(t_j)}\prod_{k \neq j}\mathcal{V}_k(t_k)]|\Omega \rangle \, .
  \label{eq:ampeps}
\end{multline}
We have freedom in choosing the normalisation constant of the soft vertex operator to match the target space gravitational coupling, which amounts to fixing  
\[
\mathcal{N}_s=-i\sqrt{32 \pi G_N} \,,
\]
and we obtain  
\begin{multline}
    \mathcal{A}(N_s, v^{I}_s, h_{s};N_1, v^{I}_1, \theta_1;...; N_n, v^{I}_n, \theta_n)\\
    =-2\sqrt{32\pi G_N}\,h_{IJ}\sum_j\eta_s \eta_j\frac{N_j}{N_s}\frac{v^I_{sj} v^J_{sj}}{v^2_{sj}}\mathcal{A}(N_1, v^{I}_1, \theta_1;...; N_n, v^{I}_n, \theta_n) \, ,
    \label{eq:leadingsoft}
\end{multline}
which is the statement of the leading soft theorem, as anticipated. Note that we restored the sum over all hard particles. The sign factors $\eta_s$ and $\eta_j$ appear for the different cases of \textit{outgoing} and \textit{incoming} $s$ and $j$ particles (recall that $\eta_i=\pm 1$ if the particle is outgoing or incoming, respectively). We can, therefore, identify the leading soft factor in (\ref{eq:softtheorem}) as
\begin{equation}
    S^{(-1)}=-2\sqrt{32\pi G_N}\,h_{IJ}\sum_j\eta_s \eta_j\frac{N_j}{N_s}\frac{v^I_{sj} v^J_{sj}}{v^2_{sj}} \,.
\end{equation}
The remaining combinations for external contractions do not modify this conclusion. The computation of the terms (\ref{eq:1hardcontr}) and (\ref{eq:2hardcontr}) proceeds analogously, and we have pointed out how these terms provide an $O(\frac{1}{N^2})$ contribution with respect to the leading order, which is subleading in the soft limit and is, therefore, negligible. 
\section{Subleading term}
\label{sec:subleading}
In this section, we derive the subleading term of the soft theorem, which scales as $N^0$ in the large $N$ limit. 
The leading term of the soft theorem is encoded in the first interaction term of the effective field theory (\ref{eq:effective}); the subleading term requires the inclusion of order $\frac{1}{r^8}$ terms in the effective interaction Lagrangian. It was shown in \cite{IV_1999} that these consist of an angular momentum part $\mathcal{L}_{ang}$ and a spin part $\mathcal{L}_{spin}$. The subleading soft factor $S^{(0)}$ in (\ref{eq:softtheorem}) receives contributions from both. 
We will analyse the two cases separately.
\subsection{Orbital Angular Momentum}
The orbital angular momentum term in (\ref{eq:effective}) (after canonically normalising the kinetic term) takes the form
\begin{equation}
    \mathcal{L}_{ang}=-\frac{7\lambda(u)}{4r^2(u)} r^{I}(u)\Bigl(\sqrt{\frac{R}{N_s}}X_s^I(u)-\sqrt{\frac{R}{N_j}}X_j^I(u)\Bigr)\biggl[\Bigl(\sqrt{\frac{R}{N_s}}\Dot{X}_s^J(u)-\sqrt{\frac{R}{N_j}}\Dot{X}_j^J(u)\Bigr)^2\biggr]^2 \,.
    \label{eq:intaddorb}
\end{equation}
In the auxiliary field formalism, the effective two-particle Lagrangian at order $o(\frac{1}{r^8})$ is 
\begin{multline}
    \mathcal{L}=\frac{1}{2}(\Dot{X}^I_s)^2+\frac{1}{2}(\Dot{X}^I_j)^2+\frac{\epsilon^{2} \lambda}{4}\Dot{\sigma}^2+\frac{\lambda}{4}\biggl(1+\frac{7}{r^2} r^I\Bigl(\sqrt{\frac{R}{N_s}}X_s^I-\sqrt{\frac{R}{N_j}}X_j^I\Bigr)\biggr)\sigma^2 \\
    -\frac{\lambda}{2}\biggl(1+\frac{7}{r^2} r^I\Bigl(\sqrt{\frac{R}{N_s}}X_s^I-\sqrt{\frac{R}{N_j}}X_j^I\Bigr)\biggr)\sigma \Bigl(\sqrt{\frac{R}{N_s}}\Dot{X}_s^I-\sqrt{\frac{R}{N_j}}\Dot{X}_j^I\Bigr)^2+\mathcal{L}_{spin}.
    \label{eq:interang}
\end{multline}
Note that the $\sigma^2$ term also receives an additional contribution. However, at this order in the EFT expansion, 
when it is reduced to order $\sigma$ by performing Wick's contraction with interaction term insertions, this simply reproduces the interaction term in (\ref{eq:interang}) and does not generate new terms. However, the angular momentum interaction term itself picks up a factor $1/2$.

If we recall that $\lambda \propto \frac{1}{r^7}$, then we can write
\begin{equation}
    \frac{r^I}{r^9}=-\frac{1}{7}\frac{\partial}{\partial \sqrt{\frac{R}{N_s}}x_s^I(u)}\Bigl(\frac{1}{r^7}\Bigr)=-\frac{1}{7}\sqrt{\frac{N_s}{R}}\frac{\partial}{\partial x_s^I(u)}\Bigl(\frac{1}{r^7}\Bigr) \, ,
    \label{eq:X1der}
\end{equation}
where $x_s^I(u)$ is the particle coordinate. When computing this derivative by parts, we can throw away the total derivative term, since the final amplitude will not depend on $x_s$ at this order of the soft expansion. Then (\ref{eq:X1der}) can be written as
\begin{equation*}
   \frac{r^I}{r^9}= \frac{1}{7}\sqrt{\frac{N_s}{R}}\Bigl(\frac{1}{r^7}\Bigr)\frac{\partial}{\partial x_s^I(u)}=\frac{1}{7}i\sqrt{\frac{N_s}{R}}\Bigl(\frac{1}{r^7}\Bigr)\dot{X}_s^I(u)
\end{equation*}
since, in the asymptotic region, the conjugate momentum to the position operator is the velocity operator. Therefore, the angular momentum interaction term in (\ref{eq:interang}) becomes
\begin{multline}
    \mathcal{L}^{\sigma}_{ang}=-\frac{i}{4}\lambda(u)\dot{X}_s^I(u)\sqrt{\frac{N_s}{R}}\Bigl(\sqrt{\frac{R}{N_s}}X_s^I(u)-\sqrt{\frac{R}{N_j}}X_j^I(u)\Bigr)\\
   \times \Bigl(\sqrt{\frac{R}{N_s}}\Dot{X}_s^J(u)-\sqrt{\frac{R}{N_j}}\Dot{X}_j^J(u)\Bigr)^2 \sigma(u) \, .
    \label{eq:angeff}
\end{multline}

We only consider the case in which (\ref{eq:angeff}) is inserted once in the amplitude, since multiple insertions lead to a contribution at higher order in the large $N$ expansion. 
The structure of internal contractions is, then, analogous to what we discussed for the leading term. The resulting interaction operator is
\begin{equation}
\label{eq:OresAng}
   O^{ang}_{res}=\frac{i}{2}\int du \,\dot{X}_s^I(u)\sqrt{\frac{N_s}{R}}\Bigl(\sqrt{\frac{R}{N_s}}X_s^I(u)-\sqrt{\frac{R}{N_j}}X_j^I(u)\Bigr)\sum_{n=1}^{\infty}\,\frac{1}{n!}\Tilde{C}_{n}(u) \,O_{res}
\end{equation}
where $\Tilde{C}_n(u)$ is the integrand of (\ref{eq:chaininteraction}). Any chain operator has two ends, as explained in section \ref{sec:leading}. We say that the chain operator $C_n$ is contracted with a velocity operator when one end is contracted with that velocity and the other end is contracted with the exponent. Moreover, we say that $C_n$ is contracted with the exponent when both ends are. The bulk of the computation is analogous to the leading term derivation. We will not, therefore, go through all the details again and will limit ourselves to an analysis of the important features of the proof for the subleading theorem. The possible options for external contractions are:
\begin{itemize}
    \item One of the soft operators is contracted with the explicit soft velocity in (\ref{eq:OresAng}), no hard operator is contracted:
    \begin{multline}
 \mathcal{A}_{sub}^{(1)}(N_s, v^{I}_s, h_s;N_1, v^{I}_1, \theta_1;...; N_n, v^{I}_n, \theta_n)\\
   =\mathcal{N}_s^*\mathcal{N}_j^*\int d\mu_{t_s}\prod_{i=1}^n \int d\mu_{t_i} \int du \,h_{IJ} \langle \Omega| T[\underline{\dot{X}^{I}_s(t_s)}\, \underline{\underline{\dot{X}^{J}_s(t_s)e^{-i\Bigl( \sqrt{\frac{2N_s}{R}}v_s \cdot X_s(t_s)+\sqrt{\frac{2N_j}{R}}v_j \cdot X_j(t_j)\Bigr)}}}\\ 
       \times  \frac{i}{2}\underline{\dot{X}_s^K(u)}\sqrt{\frac{N_s}{R}}\Bigl(\sqrt{\frac{R}{N_s}}X_s^K(u)-\sqrt{\frac{R}{N_j}}X_j^K(u)\Bigr)\sum_{n=1}^{\infty}\,\frac{1}{n!}\underline{\underline{\tilde{C}_{n}(u)\, O_{res}} }\\
       \times A_{I_1 I_2}\dot{X}_{j}^{I_1}(t_j) \dot{X}_j^{I_2}(t_j)\mathcal{V}^{*}_1(t_1)...\mathcal{V}_n(t_n)] |\Omega \rangle
        \label{eq:sub1}
\end{multline}
this will contribute to the subleading soft theorem;
\item No soft operator is contracted with the explicit soft velocity in (\ref{eq:OresAng}), no hard operator is contracted:
\begin{multline}
        \mathcal{A}_{sub}^{(2)}(N_s, v^{I}_s, h_s;N_1, v^{I}_1, \theta_1;...; N_n, v^{I}_n, \theta_n)\\
   =\mathcal{N}_s^*\mathcal{N}_j^*\int d\mu_{t_s}\prod_{i=1}^n \int d\mu_{t_i} \int du \,h_{IJ} \langle \Omega|T[\underline{\dot{X}_s^{I}(t_s)\dot{X}_s^{J}(t_s)}\,\underline{\underline{{e^{-i\Bigl( \sqrt{\frac{2N_s}{R}}v_s \cdot X_s(t_s)+\sqrt{\frac{2N_j}{R}}v_j \cdot X_j(t_j)\Bigr)}}}}\\ 
        \times  \frac{i}{2}\underline{\underline{\dot{X}_s^K(u)}}\sqrt{\frac{N_s}{R}}\Bigl(\sqrt{\frac{R}{N_s}}X_s^K(u)-\sqrt{\frac{R}{N_j}}X_j^K(u)\Bigr)\sum_{n=1}^{\infty}\,\frac{1}{n!}\underline{\tilde{C}_{n}(u)\, O_{res}}\\
         \times A_{I_1 I_2}\dot{X}_{j}^{I_1}(t_j) \dot{X}_j^{I_2}(t_j)\mathcal{V}^{*}_1(t_1)...\mathcal{V}_n(t_n)] |\Omega\rangle
        \label{eq:sub2}
        \end{multline}
        this term will also contribute to the subleading soft theorem;
        \item One of the soft operators is contracted with index $K$ velocity, and one of the hard operators is contracted:
        \begin{multline}
       \mathcal{A}_{sub}^{(3)}(N_s, v^{I}_s, h_s;N_1, v^{I}_1, \theta_1;...; N_n, v^{I}_n, \theta_n)\\
   =\mathcal{N}_s^*\mathcal{N}_j^*\int d\mu_{t_s}\prod_{i=1}^n \int d\mu_{t_i}  \int du \,h_{IJ} \langle \Omega| T[\underline{\dot{X}^{I}_s(t_s)}\,\underbrace{\dot{X}^{J}_j(t_s)e^{-i\Bigl( \sqrt{\frac{2N_s}{R}}v_s \cdot X_s(t_s)+\sqrt{\frac{2N_j}{R}}v_j \cdot X_j(t_j)\Bigr)}}\\ 
         \times \frac{i}{2}\underline{\dot{X}_s^K(u)}\sqrt{\frac{N_s}{R}}\Bigl(\sqrt{\frac{R}{N_s}}X_s^K(u)-\sqrt{\frac{R}{N_j}}X_j^K(u)\Bigr)\sum_{n=1}^{\infty}\,\frac{1}{n!}\underline{\underline{\tilde{C}_{n}(u)}}\underbrace{O_{res}}\\
       \times  A_{I_1 I_2}\underline{\underline{\dot{X}_{j}^{I_1}(t_j)}}\dot{X}^{I_2}_j(t_j)\mathcal{V}^{*}_1(t_1)...\mathcal{V}_n(t_n)] |\Omega\rangle\,,
        \end{multline}
        where the under-brace is used to distinguish between different Wick contractions. This term gives a contribution of order $O(\frac{1}{N^2})$ with respect to the subleading order, so it does not contribute to the soft theorem;
        \item No soft operator is contracted with index $K$ velocity, one hard operator is contracted:
        \begin{multline}
        \mathcal{A}_{sub}^{(4)}(N_s, v^{I}_s, h_s;N_1, v^{I}_1, \theta_1;...; N_n, v^{I}_n, \theta_n)\\
   =\mathcal{N}_s^*\mathcal{N}_j^*\int d\mu_{t_s}\prod_{i=1}^n \int d\mu_{t_i}\int du \,h_{IJ} \langle \Omega| T[\underline{\dot{X}^{I}_s(t_s)}\,\underline{\underline{\dot{X}^{J}_s(t_s)}}\underbrace{e^{-i\Bigl( \sqrt{\frac{2N_s}{R}}v_s \cdot X_s(t_s)+\sqrt{\frac{2N_j}{R}}v_j \cdot X_j(t_j)\Bigr)}}\\ 
        \times \frac{i}{2}\underbrace{\dot{X}_s^K(u)}\sqrt{\frac{N_s}{R}}\Bigl(\sqrt{\frac{R}{N_s}}X_s^K(u)-\sqrt{\frac{R}{N_j}}X_j^K(u)\Bigr)\sum_{n=1}^{\infty}\,\frac{1}{n!}\underline{\tilde{C}_{n}(u)}\, \underline{\underline{O_{res}}}\\
      \times A_{I_1 I_2}\underline{\dot{X}_{j}^{I_1}(t_j)}\dot{X}^{I_2}_j(t_j) \mathcal{V}^{*}_1(t_1)...\mathcal{V}_n(t_n)] | \Omega\rangle
        \end{multline}
    this term does not contribute for the same reason as the previous;
        \item Both hard operators are contracted:
        \begin{multline}
        \mathcal{A}_{sub}^{(5)}(N_s, v^{I}_s, h_s;N_1, v^{I}_1, \theta_1;...; N_n, v^{I}_n, \theta_n)\\
   =\mathcal{N}_s^*\mathcal{N}_j^*\int d\mu_{t_s}\prod_{i=1}^n \int d\mu_{t_i}\int du \,h_{IJ} \langle \Omega| T[\underline{\dot{X}_s^{I}(t_s)\dot{X}_s^{J}(t_s)}\,\underline{\underline{e^{-i\Bigl( \sqrt{\frac{2N_s}{R}}v_s \cdot X_s(t_s)+\sqrt{\frac{2N_j}{R}}v_j \cdot X_j(t_j)\Bigr)}}}\\ 
        \times \frac{i}{2}\underline{\underline{\dot{X}_s^K(u)}}\sqrt{\frac{N_s}{R}}\Bigl(\sqrt{\frac{R}{N_s}}X_s^K(u)-\sqrt{\frac{R}{N_j}}X_j^K(u)\Bigr)\sum_{n=1}^{\infty}\,\frac{1}{n!}\underline{\tilde{C}_{n}(u)\, O_{res}}\\
        \times A_{I_1 I_2}\underline{\dot{X}_{j}^{I_1}(t_j) \dot{X}_j^{I_2}(t_j)} \mathcal{V}^{*}_1(t_1)...\mathcal{V}_n(t_n)]|\Omega \rangle
        \end{multline}
        this term is also $O(\frac{1}{N^2})$ with respect to the subleading order, so it does not contribute to the soft theorem.
\end{itemize}
 We see that the only relevant contributions come from (\ref{eq:sub1}) and (\ref{eq:sub2}). Note that we did not explicitly list any option in which the factor $\Bigl(\sqrt{\frac{R}{N_s}}X_s^I(u)-\sqrt{\frac{R}{N_j}}X_j^I(u)\Bigr)$ is contracted. The reason is that the contractions involving the soft operator $X_s^I(u)$ give a total derivative term which vanishes due to $\delta$ functions coming from other contractions. Moreover, contractions with prefactors which involve a single chain yield potentially divergent terms, which are set to zero by dimensional regularisation, as explained in the appendix \ref{sec:app}. As for the hard operator $X_j^I(u)$, after external contractions, this will be evaluated at $t_s$ due to the $\delta(u-t_s)$ factor that comes out of propagators. Then it can be integrated by parts, yielding a factor proportional to $X_j^I(t_j)$ plus a term which vanishes at this order in the $\frac{1}{N}$ expansion.
We also note that $X_j^I(t_j)$ can be extracted from the hard exponent as
\begin{equation}
    X_j^I(t_j)=i\sqrt{\frac{R}{2N_j}}\frac{\partial}{\partial v_j^I} e^{-i\sqrt{\frac{2N_j}{R}}v_j \cdot X_j(t_j)}.
    \label{eq:der}
\end{equation}
All in all, considering the result of Wick's contractions for $\mathcal{A}_{sub}^{(1)}(N_s, v^{I}_s, h_s;N_1, v^{I}_1, \theta_1;...; N_n, v^{I}_n, \theta_n) $ in (\ref{eq:sub1}), proceeding through steps that are analogous to those for the leading term, the resulting S-matrix element reads (restoring $\eta_s$, $\eta_j$ factors)
\begin{multline}
\mathcal{A}_{sub}^{(1)}(N_s, v^{I}_s, h_{I J};N_1, v^{I}_1, \theta_1;...; N_n, v^{I}_n, \theta_n)\\
=\frac{1}{4}\sqrt{32\pi G_N}\,h_{IJ}\sum_j\eta_s \eta_j \bigl(v^I_{sj}\partial_{v_j^J}+v^J_{sj}\partial_{v_j^I}\bigr)\mathcal{A}(N_1, v^{I}_1, \theta_1;...; N_n, v^{I}_n, \theta_n)\,.
    \label{eq:part1ang}
\end{multline}
The same considerations apply to the contribution (\ref{eq:sub2}). The corresponding matrix element is
\begin{multline}
    \mathcal{A}_{sub}^{(2)}(N_s, v^{I}_s, h_{I J};N_1, v^{I}_1, \theta_1;...; N_n, v^{I}_n, \theta_n)\\
    =-\sqrt{32\pi G_N}\, h_{IJ}\sum_j \eta_s \eta_j \frac{v^I_{sj}v^J_{sj}}{v^2_{sj}}v_{s}^K\partial_{v^K_j}
   \mathcal{A}(N_1, v^{I}_1, \theta_1;...; N_n, v^{I}_n, \theta_n)\\
    =-\sqrt{32\pi G_N}\, h_{IJ}\sum_j\eta_s \eta_j \Bigl(\frac{v^I_{sj}v^J_{sj}}{v^2_{sj}}v_{sj}^K\partial_{v^K_j}+2\frac{N_j}{R} \frac{v^I_{sj}v^J_{sj}}{v^2_{sj}}\partial_{\frac{N_j}{R}}\Bigr)\mathcal{A}(N_1, v^{I}_1, \theta_1;...; N_n, v^{I}_n, \theta_n)
    \\
    =-\sqrt{32\pi G_N}\,h_{IJ}\sum_j\eta_s \eta_j \Bigl(\frac{v^I_{sj}v^J_{sj}}{v^2_{sj}}v_{sj}^K\partial_{v^K_j}+2\frac{N_j}{R} \frac{v^I_{sj}v^J_{sj}}{v^2_{sj}}\partial_{\frac{N_j}{R}}\Bigr)\mathcal{A}(N_1, v^{I}_1, \theta_1;...; N_n, v^{I}_n, \theta_n) \, ,
    \label{eq:part2ang}
\end{multline}
where we used
\begin{equation}
     v_j \cdot \partial_{v_j} e^{-i\sqrt{\frac{2N_j}{R}}v_j\cdot X_j(t)}=2\frac{N_j}{R}\partial_{\frac{N_j}{R}}e^{-i\sqrt{\frac{2N_j}{R}}v_j\cdot X_j(t)}
\end{equation}
in the second equality. Equations (\ref{eq:part1ang}) and (\ref{eq:part2ang}) represent the orbital momentum contribution to the soft factorisation.
\subsection{Spin}
After canonically normalising the kinetic term, the spin contribution to (\ref{eq:effective}) at order $\frac{1}{r^8}$ is \cite{IV_1999}
\begin{multline}
    \mathcal{L}_{\text{spin}}=-\frac{R}{N_j}\frac{7\lambda(u)}{2 r^2(u)}r^I(u)\Bigl(\sqrt{\frac{R}{N_s}}\Dot{X}^K_s(u)-\sqrt{\frac{R}{N_j}}\Dot{X}^K_j(u)\Bigr)^2 \\
    \times S_j^{IJ}\Bigl(\sqrt{\frac{R}{N_s}}\Dot{X}^J_s(u)-\sqrt{\frac{R}{N_j}}\Dot{X}^J_j(u)\Bigr) \, .
    \label{eq:Spinterm}
\end{multline}
Here $S^{IJ}_j$ stands for the BFSS spin generator acting on the $j$-th particle, defined as
\begin{equation}
    S^{IJ}_j=\frac{1}{8R}Tr(\psi^{\alpha} \Gamma^{[IJ]}_{\alpha \beta} \psi^{\beta}) \, ,
\end{equation}
where $\psi$ are BFSS fermion fields and $\Gamma^{[IJ]}$ are Lorentz generators built from gamma matrices. We will not rephrase (\ref{eq:Spinterm}) with an auxiliary field. As in the case of the orbital angular momentum, we can rewrite (\ref{eq:Spinterm}) to obtain
\begin{multline}
   \mathcal{L}_{\text{spin}}= -i\frac{R}{2N_j} \lambda(u)\sqrt{\frac{N_s}{R}} \Bigl(\sqrt{\frac{R}{N_s}}\Dot{X}^K_s(u)-\sqrt{\frac{R}{N_j}}\Dot{X}^K_j(u)\Bigr)^2 \\
   \times S_j^{IJ}\Dot{X}^I_s(u)\Bigl(\sqrt{\frac{R}{N_s}}\Dot{X}^J_s(u)-\sqrt{\frac{R}{N_j}}\Dot{X}^J_j(u)\Bigr)
    \label{eq:Spintermrep}
\end{multline}
by using the argument around (\ref{eq:X1der}). As usual, internal contractions dress the interaction term. Since we are not working in the auxiliary field formalism for this term, there is still a self-interacting contribution which produces a non-vacuum-vacuum diagram. It follows that, for this term, we have two possible options for internal contractions:
\begin{itemize}
    \item Self-interacting contribution:
    first we contract two of the velocities at time $u$ among themselves, and then we construct chains. This evaluates to
    \begin{multline}
    O^{s}_{res}=-i\int du \,\sqrt{\frac{N_s}{R}}\Bigl(\frac{R}{N_j}\Bigr)\frac{\lambda(u)}{\epsilon}\bigl(1+F[\sigma(u)]\bigr)\\
    S_j^{IJ}\dot{X}^I_s(u)\Bigl(\sqrt{\frac{R}{N_s}}\dot{X}^J_s(u)-\sqrt{\frac{R}{N_j}}\dot{X}^J_j(u)\Bigr)
     O_{res} \, ;
\end{multline}
\item Non-self-interacting contribution: two chains are attached to the new interaction operator. This gives
\begin{multline}
    O^{ns}_{res}=-i\frac{N_j}{R}\sqrt{\frac{N_s}{R}}S_j^{IJ}\int du \,\lambda(u)\bigl(1+F[\sigma(u)]\bigr)^2\Dot{X}^I_s(u)\\
    \times \Bigl(\sqrt{\frac{R}{N_s}}\Dot{X}^J_s(u)-\sqrt{\frac{R}{N_j}}\Dot{X}^J_j(u)\Bigr)\Bigl(\sqrt{\frac{R}{N_s}}\Dot{X}^K_s(u)-\sqrt{\frac{R}{N_j}}\Dot{X}^K_j(u)\Bigr)^2
     O_{res} \, .
    \end{multline}
\end{itemize}
Each of these options is associated with a unique non-vanishing external contraction:
\begin{itemize}
    \item Self-interacting contribution:
    \begin{multline}
            \mathcal{A}^{(1)}_{spin}(N_s, v^{I}_s, h_s;N_1, v^{I}_1, \theta_1;...; N_n, v^{I}_n, \theta_n)\\
   =\mathcal{N}_s^*\mathcal{N}_j^*\int d\mu_{t_s}\prod_{i=1}^n \int d\mu_{t_i}h_{IJ}\langle \Omega| T[\underline{\dot{X}_s^{I}(t_s)\dot{X}_s^{J}(t_s)e^{-i\Bigl( \sqrt{\frac{2N_s}{R}}v_s \cdot X_s(t_s)+\sqrt{\frac{2N_j}{R}}v_j \cdot X_j(t_j)\Bigr)}} \, \underline{O^{s}_{res}}\\[8pt] \times A_{I_1 I_2}\dot{X}_{j}^{I_1}(t_j) \dot{X}_j^{I_2}(t_j)\mathcal{V}^{*}_1(t_1)...\mathcal{V}_n(t_n)]|\Omega \rangle \\[8pt]
    \end{multline}
    which gives a contribution $O(\frac{1}{N^2})$ with respect to the subleading term;
    \item Non-self-interacting contribution:
     \begin{multline}
        \mathcal{A}^{(2)}_{spin}(N_s, v^{I}_s, h_s;N_1, v^{I}_1, \theta_1;...; N_n, v^{I}_n, \theta_n)\\
   =\mathcal{N}_s^*\mathcal{N}_j^*\int d\mu_{t_s}\prod_{i=1}^n \int d\mu_{t_i}h_{IJ}\langle \Omega | T[\underline{\dot{X}_s^{I}(t_s)\dot{X}_s^{J}(t_s)e^{-i\Bigl( \sqrt{\frac{2N_s}{R}}v_s \cdot X_s(t_s)+\sqrt{\frac{2N_j}{R}}v_j \cdot X_j(t_j)\Bigr)}} \, \underline{O^{ns}_{res}} \\[8pt] \times A_{I_1 I_2}\dot{X}_{j}^{I_1}(t_j) \dot{X}_j^{I_2}(t_j)\mathcal{V}^{*}_1(t_1)...\mathcal{V}_n(t_n)]| \Omega \rangle\\[8pt]
        =2i\sqrt{32\pi G_N}\frac{h_{IJ}v^I_{sj}}{v^2_{sj}}S_j^{JK}v^K_{j}\prod_{i=1}^n \int d\mu_{t_i}\langle \Omega| T[\mathcal{V}^{*}_j(t_j)\mathcal{V}^{*}_1(t_1)...\mathcal{V}_n(t_n)]| \Omega\rangle \, .
    \end{multline}
\end{itemize}
The contribution of other contractions vanishes, analogously to the case of the orbital angular momentum. Out of those, only the second contributes to the subleading order of the soft expansion. It follows that the spin contribution to the soft theorem can be written as (restoring $\eta_s$, $\eta_j$ factors)
\begin{multline}
    \mathcal{A}_{spin}(N_s,v_s,h_s;N_1,v_1,\theta_1,...,N_n,v_n,\theta_n)\\[5pt]
    =2i\sqrt{32\pi G_N}\sum_j \eta_s \eta_j \frac{h_{IJ}v^I_{sj}}{v^2_{sj}}S_j^{JK}v^K_{j}\mathcal{A}(N_1,v_1,\theta_1,...,N_n,v_n,\theta_n)\, ,
\end{multline}
which represents the spin contribution to the subleading soft factor.

We can collect the orbital angular momentum term and the spin term together and summarise the subleading contribution to the soft theorem. This reads
\begin{multline}
S^{(0)}=\frac{1}{4}\sqrt{32\pi G_N}\sum_{j=1}^n \eta_s \eta_j h_{IJ}\Bigl(\bigl(v^I_{sj}\partial_{v_j^J}+v^J_{sj}\partial_{v_j^I}\bigr)\\
-4\,\frac{v^I_{sj}v^J_{sj}}{v^2_{sj}}v_{sj}^K\partial_{v^K_j}
-8\frac{N_j}{R} \frac{v^I_{sj}v^J_{sj}}{v^2_{sj}}\partial_{\frac{N_j}{R}}+8i\frac{v^I_{sj}}{v^2_{sj}}S_j^{JK}v^K_{j}\Bigr)\label{eqn:subleadingS}
\end{multline}
such that
\begin{multline}
    \mathcal{A}(N_s,v_s,h_s;N_1,v_1,\theta_1,...,N_n,v_n,\theta_n) \\
=\Bigl(S^{(-1)}+S^{(0)}+...\Bigr)\mathcal{A}(N_1,v_1,\theta_1,...,N_n,v_n,\theta_n) \, ,
\end{multline}
which reproduces the result in \cite{IR}, corresponding to the statement of the leading and subleading soft theorem for the BFSS theory. 

As a final remark, we note that, in the EFT approach adopted in this paper, we have truncated all higher-order terms in the $O(\frac{1}{r})$ expansion from the Lagrangian (\ref{eq:effective}), as they give a suppressed contribution in the large distance regime of our computation. These terms include the supersymmetric completions of the $O(\frac{\dot{X}^4}{r^7})$ term with a higher number of fermionic fields (see \cite{Sethi:1999}), and terms of order $O(\frac{\dot{X}^6}{r^{14}})$ or higher. Interestingly, as far as powers of $N$ are concerned, the latter is the only term which would mix up with the subleading order of the soft expansion. All the others will give contributions which are also suppressed by powers of $O(\frac{1}{N})$ after considering their contractions with external vertex operators. 
\section{Conclusion}
\label{sec:conclusion}
In this paper, we derived a soft theorem for the BFSS theory by considering scattering amplitudes in the effective theory describing two-particle interactions between $D_0$ brane bound states. In particular, the objects whose amplitudes enjoy a soft factorisation are vertex-like operators with the correct quantum numbers to describe supergravitons in the target space. We showed that the lowest order in the large distance expansion, corresponding to $\sim \frac{1}{r^7}$ encodes the leading factor of the soft theorem, whereas terms at order $\sim \frac{1}{r^8}$ give rise to the subleading factor. The calculation requires the introduction of a term in the Lagrangian which is only relevant in the UV, after which the theory is super-renormalisable in dimensional regularisation. The soft correction to the correlation function is expanded in powers of $N$, and the leading $O(N)$ and subleading $O(N^0)$ soft factors are obtained. The result has precisely the form required by the soft theorem in BFSS. The expressions we obtain are, in fact, consistent with the ones predicted in \cite{IR} (up to a matching of the overall constant multiplying the amplitude), which are obtained by applying the BFSS duality dictionary to translate from the soft graviton theorem in 11-dimensional supergravity. In light of the connection between soft theorem and asymptotic symmetries, we conclude that, in the regime of validity of the effective field theory picture, our analysis confirms the realisation of the full 11-dimensional Lorentz symmetry, along with the presence of an infinite-dimensional asymptotic symmetry group. 

A few questions remain open. First, it would be interesting to study how the soft theorem emerges from the full matrix description of BFSS theory. This would shed light on the Matrix-theoretical structure of the theory and would, perhaps, provide important insights on concepts such as emergent space-time and emergent symmetries from the Matrix Quantum Mechanics description of M-theory, possibly away from the supergravity limit. It would also be interesting to investigate the realisation of soft theorems for the gravitino and three-form field in BFSS and their connection to asymptotic symmetries. A full understanding of soft theorems from a Matrix Quantum Mechanics perspective would help to answer these questions. 
\acknowledgments 
We would like to thank Ludovic Fraser-Taliente and Ryan Barouki for their useful discussions. We thank Savdeep Sethi for useful comments on the first version of the manuscript. DL is supported by STFC studentship ST/X508664/1.
\appendix
\section{UV finiteness of the effective theory}
\label{sec:app}
Here we show that the theory defined by the Lagrangian (\ref{eq:finallag}) does not feature UV divergences, as long as we keep the regulator $\epsilon$ to be non-zero.

Although we worked in position space in the main text, it will be convenient to express the arguments of the present appendix in momentum space. We will be concerned with diagrams involving propagators between two velocity operators $\Dot{X}$ and between two $\sigma$ fields. Note that the momentum space version of (\ref{eq:twovel}) and (\ref{eq:twoaux}) reads
\begin{equation}
\langle \sqrt{\lambda}\sigma(p) \sqrt{\lambda}\sigma(0) \rangle =\frac{i}{\epsilon^2 p^2+1}    
\end{equation}
and 
\begin{equation}
    \langle \Dot{X}(p) \Dot{X}(0) \rangle =i
\end{equation}
respectively. It follows that the degree of divergence $\omega$ of a given diagram is given by 
\begin{equation}
    \omega = L-2 n_{\sigma}
\end{equation}
where $L$ is the number of loops and $n_{\sigma}$ is the number of internal $\sigma$ lines. This is also equal to
\begin{equation}
    \omega=-\frac{v}{2}+\frac{l_{\sigma}}{2}-\frac{l_{\Dot{X}}}{2}+1 \, ,
\end{equation}
where $v$ is the number of vertices, $l_{\sigma}$ is the number of external $\sigma$ lines and $l_{\Dot{X}}$ is the number of external $\Dot{X}$ lines. Let us analyse the degree of divergence case by case:
\begin{itemize}
    \item No external $\sigma$ legs and $l_{\Dot{X}}\geq 2$
    \begin{equation}
        l_{\sigma}=0, \quad l_{\Dot{X}}\geq 2 \quad  \Rightarrow \quad \omega <0
    \end{equation}
    These diagrams are convergent.
    \item No external $\Dot{X}$ legs, $l_{\Dot{X}}=0$, in this case 
    \begin{equation}
        L=1+n_{\sigma} \quad \Rightarrow \quad \omega = 1 - n_{\sigma} \, .
    \end{equation}
    then for $n_{\sigma}=0$ the diagram is linearly divergent. However, the divergence is proportional to $\int dp$, which vanishes in dimensional regularisation. The corresponding Feynman diagram is shown in Figure \ref{fig:Circle}.
    \begin{figure}[H]
\centering
\includegraphics[width=0.3\textwidth]{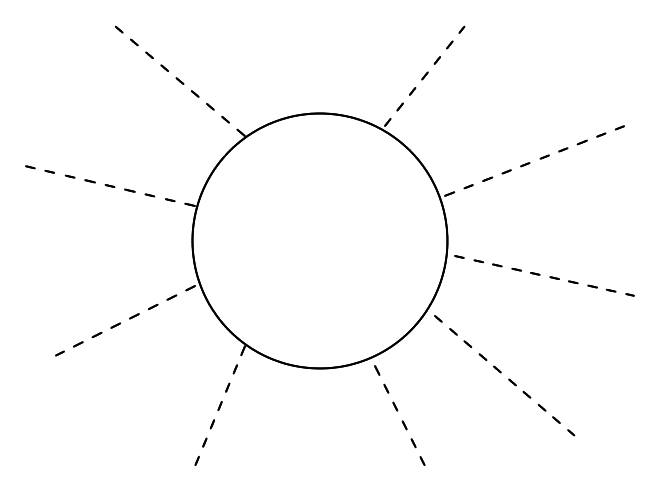}
\caption{Linearly divergent diagram with $l_{\Dot{X}}=0$ and $n_{\sigma}=0$. This vanishes in dimensional regularisation.}
\label{fig:Circle}
\end{figure} For $n_{\sigma}=1$, $\omega=0$, and we have an overlapping log divergence which takes the form $\int dr dp \,\frac{1}{\epsilon^2 p^2+1}$, which again vanishes due to $\int dr$ in dimensional regularisation. The corresponding Feynman diagram is shown in Figure \ref{fig:Square}.
    \begin{figure}[h]
\centering
\includegraphics[width=0.3\textwidth]{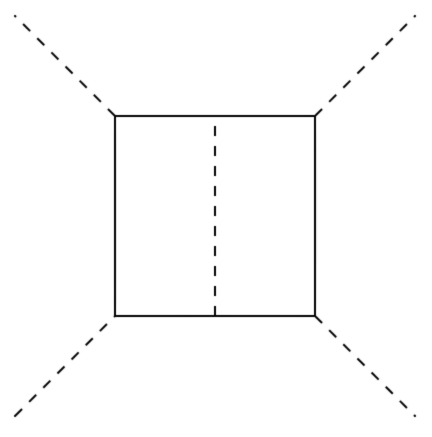}
\caption{Overlapping log divergent diagram with $l_{\Dot{X}}=0$ and $n_{\sigma}=1$. This vanishes in dimensional regularisation.}
\label{fig:Square}
\end{figure} 
 
Indeed, if we go to higher values of $n_{\sigma}$, $\omega$ becomes negative and the corresponding diagrams are convergent;
    \item Two external $\Dot{X}$ legs, $l_{\Dot{X}}=2$, in this case
    \begin{equation}
        L=n_{\sigma} \quad \Rightarrow \quad \omega = - n_{\sigma} \, .
    \end{equation}
    Then the only possible divergence arises when $n_{\sigma}=0$, then $\omega=0$. However, there is no Feynman diagram corresponding to this situation. 
    \item $l_{\Dot{X}}\geq 4$, then $\omega<0$ and the corresponding diagrams are convergent.
\end{itemize}
To summarise, due to the fact that the theory is one-dimensional, the structure of divergences is very simple, and not only is the theory super-renormalizable, but all possible divergences cancel out.

\bibliographystyle{JHEP}
\bibliography{references.bib}

@article{BFSS,
    author = "Banks, Tom and Fischler, W. and Shenker, S. H. and Susskind, Leonard",
    title = "{M theory as a matrix model: A conjecture}",
    eprint = "hep-th/9610043",
    archivePrefix = "arXiv",
    reportNumber = "RU-96-95, SU-ITP-96-12, UTTG-13-96",
    doi = "10.1201/9781482268737-37",
    journal = "Phys. Rev. D",
    volume = "55",
    pages = "5112--5128",
    year = "1997"
}

@article{susskind1997conjecture,
    author = "Susskind, Leonard",
    title = "{Another conjecture about M(atrix) theory}",
    eprint = "hep-th/9704080",
    archivePrefix = "arXiv",
    reportNumber = "SU-ITP-97-11",
    month = "4",
    year = "1997"
}

@article{IR,
    author = "Tropper, Adam and Wang, Tianli",
    title = "{Lorentz symmetry and {IR} structure of the {BFSS} matrix model}",
    eprint = "2303.14200",
    archivePrefix = "arXiv",
    primaryClass = "hep-th",
    doi = "10.1007/JHEP07(2023)150",
    journal = "JHEP",
    volume = "07",
    pages = "150",
    year = "2023"
}

@article{IV_1999,
    author = "Taylor, Washington and Van Raamsdonk, Mark",
    title = "{Supergravity currents and linearized interactions for matrix theory configurations with fermionic backgrounds}",
    eprint = "hep-th/9812239",
    archivePrefix = "arXiv",
    reportNumber = "PUPT-1828, MIT-CTP-2810",
    doi = "10.1088/1126-6708/1999/04/013",
    journal = "JHEP",
    volume = "04",
    pages = "013",
    year = "1999"
}

@article{Okawa_1999,
    author = "Okawa, Yuji and Yoneya, Tamiaki",
    title = "{Multibody interactions of D particles in supergravity and matrix theory}",
    eprint = "hep-th/9806108",
    archivePrefix = "arXiv",
    reportNumber = "UT-KOMABA-98-13",
    doi = "10.1016/S0550-3213(98)00700-7",
    journal = "Nucl. Phys. B",
    volume = "538",
    pages = "67--99",
    year = "1999"
}

@article{Okawa_1999_2,
    author = "Okawa, Yuji and Yoneya, Tamiaki",
    title = "{Equations of motion and Galilei invariance in D particle dynamics}",
    eprint = "hep-th/9808188",
    archivePrefix = "arXiv",
    reportNumber = "UT-KOMABA-98-21",
    doi = "10.1016/S0550-3213(98)00769-X",
    journal = "Nucl. Phys. B",
    volume = "541",
    pages = "163--178",
    year = "1999"
}

@article{Becker_1997,
    author = "Becker, Katrin and Becker, Melanie",
    title = "{A Two loop test of M(atrix) theory}",
    eprint = "hep-th/9705091",
    archivePrefix = "arXiv",
    reportNumber = "NSF-ITP-97-047",
    doi = "10.1016/S0550-3213(97)00518-X",
    journal = "Nucl. Phys. B",
    volume = "506",
    pages = "48--60",
    year = "1997"
}

@article{Becker_1997_2,
    author = "Becker, Katrin and Becker, Melanie and Polchinski, Joseph and Tseytlin, Arkady A.",
    title = "{Higher order graviton scattering in M(atrix) theory}",
    eprint = "hep-th/9706072",
    archivePrefix = "arXiv",
    reportNumber = "NSF-ITP-97-061, IMPERIAL-TP-96-97-49",
    doi = "10.1103/PhysRevD.56.R3174",
    journal = "Phys. Rev. D",
    volume = "56",
    pages = "R3174--R3178",
    year = "1997"
}

@article{Kabat_1998,
    author = "Kabat, Daniel N. and Taylor, Washington",
    title = "{Linearized supergravity from matrix theory}",
    eprint = "hep-th/9712185",
    archivePrefix = "arXiv",
    reportNumber = "IASSNS-HEP-97-141, PUPT-1751",
    doi = "10.1016/S0370-2693(98)00281-0",
    journal = "Phys. Lett. B",
    volume = "426",
    pages = "297--305",
    year = "1998"
}

@article{Plefka_1998,
	doi = {10.1016/s0550-3213(97)00762-1},
  
	url = {https://doi.org/10.1016%2Fs0550-3213%2897%2900762-1},
  
	year = 1998,
	month = {feb},
  
	publisher = {Elsevier {BV}
},
  
	volume = {512},
  
	number = {1-2},
  
	pages = {460--484},
  
	author = {Jan Plefka and Andrew Waldron},
  
	title = {On the quantum mechanics of {M} (atrix) theory},
  
	journal = "Nucl. Phys. B"

}

@article{Plefka2_1998,
    author = "Plefka, Jan C. and Serone, Marco and Waldron, Andrew K.",
    title = "{The Matrix theory S matrix}",
    eprint = "hep-th/9806081",
    archivePrefix = "arXiv",
    reportNumber = "NIKHEF-98-016, UVA-WINS-WISK-98-06",
    doi = "10.1103/PhysRevLett.81.2866",
    journal = "Phys. Rev. Lett.",
    volume = "81",
    pages = "2866--2869",
    year = "1998"
}

@article{Serone_2000,
    author = "Plefka, J. and Serone, M. and Waldron, A.",
    editor = "Lust, D. and Otto, H. J.",
    title = "{Matrix theory and Feynman diagrams}",
    eprint = "hep-th/9903099",
    archivePrefix = "arXiv",
    reportNumber = "AEI-104, SPIN-1999-04",
    doi = "10.1002/(SICI)1521-3978(20001)48:1/3<191::AID-PROP191>3.0.CO;2-#",
    journal = "Fortsch. Phys.",
    volume = "48",
    pages = "191--194",
    year = "2000"
}

@inproceedings{plefka1998asymptotic,
    author = "Plefka, Jan and Waldron, Andrew",
    title = "{Asymptotic supergraviton states in matrix theory}",
    booktitle = "{31st International Ahrenshoop Symposium on the Theory of Elementary Particles}",
    eprint = "hep-th/9801093",
    archivePrefix = "arXiv",
    reportNumber = "NIKHEF-98-001",
    pages = "130--136",
    month = "9",
    year = "1997"
}

@article{Helling_1999,
    author = "Helling, Robert and Plefka, Jan and Serone, Marco and Waldron, Andrew",
    title = "{Three graviton scattering in M theory}",
    eprint = "hep-th/9905183",
    archivePrefix = "arXiv",
    reportNumber = "AEI-112, UVA-WINS-WISK-99-09, NIKHEF-99-014",
    doi = "10.1016/S0550-3213(99)00451-4",
    journal = "Nucl. Phys. B",
    volume = "559",
    pages = "184--204",
    year = "1999"
}

@article{Becker_1998,
    author = "Becker, Katrin and Becker, Melanie",
    title = "{On graviton scattering amplitudes in M theory}",
    eprint = "hep-th/9712238",
    archivePrefix = "arXiv",
    reportNumber = "CALT-68-2149",
    doi = "10.1103/PhysRevD.57.6464",
    journal = "Phys. Rev. D",
    volume = "57",
    pages = "6464--6470",
    year = "1998"
}

@article{Weinberg,
  title = {Infrared Photons and Gravitons},
  author = {Weinberg, Steven},
  journal = {Phys. Rev.},
  volume = {140},
  issue = {2B},
  pages = {B516--B524},
  numpages = {0},
  year = {1965},
  month = {Oct},
  publisher = {American Physical Society},
  doi = {10.1103/PhysRev.140.B516},
  url = {https://link.aps.org/doi/10.1103/PhysRev.140.B516}

}

@article{Taylor_2001,
    author = "Taylor, Washington",
    title = "{M(atrix) Theory: Matrix Quantum Mechanics as a Fundamental Theory}",
    eprint = "hep-th/0101126",
    archivePrefix = "arXiv",
    doi = "10.1103/RevModPhys.73.419",
    journal = "Rev. Mod. Phys.",
    volume = "73",
    pages = "419--462",
    year = "2001"
}

@article{Polchinski_1999,
    author = "Polchinski, Joseph",
    editor = "Iso, S. and Kawai, H. and Natsuume, M.",
    title = "{M theory and the light cone}",
    eprint = "hep-th/9903165",
    archivePrefix = "arXiv",
    reportNumber = "NSF-ITP-99-17",
    doi = "10.1143/PTPS.134.158",
    journal = "Prog. Theor. Phys. Suppl.",
    volume = "134",
    pages = "158--170",
    year = "1999"
}

@article{ydri2018review,
    author = "Ydri, Badis",
    title = "{Review of M(atrix)-Theory, Type IIB Matrix Model and Matrix String Theory}",
    eprint = "1708.00734",
    archivePrefix = "arXiv",
    primaryClass = "hep-th",
    month = "8",
    year = "2017"
}

@article{Seiberg_1997,
	doi = {10.1103/physrevlett.79.3577},
	url = {https://doi.org/10.1103%2Fphysrevlett.79.3577},
	year = 1997,
	month = {nov},
	publisher = {American Physical Society ({APS})},
	volume = {79},
	number = {19},
	pages = {3577--3580},
	author = {Nathan Seiberg},
	title = {Why Is the Matrix Model Correct?},
	journal = {Physical Review Letters},
eprint         = "hep-th/9710009",
      archivePrefix  = "arXiv",
      primaryClass   = "hep-th"

}

@article{miller2022soft,
    author = "Miller, Noah and Strominger, Andrew and Tropper, Adam and Wang, Tianli",
    title = "{Soft gravitons in the BFSS matrix model}",
    eprint = "2208.14547",
    archivePrefix = "arXiv",
    primaryClass = "hep-th",
    doi = "10.1007/JHEP11(2023)174",
    journal = "JHEP",
    volume = "11",
    pages = "174",
    year = "2023"
}

@article{bigatti1997review,
    author = "Bigatti, D. and Susskind, Leonard",
    editor = "Baulieu, L. and Kazakov, V. and Picco, M. and Windey, Paul and Di Francesco, P. and Douglas, Michael R.",
    title = "{Review of matrix theory}",
    eprint = "hep-th/9712072",
    archivePrefix = "arXiv",
    reportNumber = "SU-ITP-97-51",
    journal = "NATO Sci. Ser. C",
    volume = "520",
    pages = "277--318",
    year = "1999"
}

@article{DANIELSSON_1996,
    author = "Danielsson, Ulf H. and Ferretti, Gabriele and Sundborg, Bo",
    title = "{D particle dynamics and bound states}",
    eprint = "hep-th/9603081",
    archivePrefix = "arXiv",
    reportNumber = "USITP-96-03, UUITP-2-96",
    doi = "10.1142/S0217751X96002492",
    journal = "Int. J. Mod. Phys. A",
    volume = "11",
    pages = "5463--5478",
    year = "1996"
}

@article{Kabat_1996,
    author = "Kabat, Daniel N. and Pouliot, Philippe",
    title = "{A Comment on zero-brane quantum mechanics}",
    eprint = "hep-th/9603127",
    archivePrefix = "arXiv",
    reportNumber = "RU-96-17",
    doi = "10.1103/PhysRevLett.77.1004",
    journal = "Phys. Rev. Lett.",
    volume = "77",
    pages = "1004--1007",
    year = "1996"
}

@article{Bachas_1996,
    author = "Bachas, C.",
    title = "{D-brane dynamics}",
    eprint = "hep-th/9511043",
    archivePrefix = "arXiv",
    reportNumber = "NSF-ITP-95-144, CPTH-S388-1195",
    doi = "10.1016/0370-2693(96)00238-9",
    journal = "Phys. Lett. B",
    volume = "374",
    pages = "37--42",
    year = "1996"
}

@article{Sethi_1998,
    author = "Sethi, Savdeep and Stern, Mark",
    title = "{D-brane bound states redux}",
    eprint = "hep-th/9705046",
    archivePrefix = "arXiv",
    reportNumber = "IASSNS-HEP-97-45, DUK-M-97-5",
    doi = "10.1007/s002200050374",
    journal = "Commun. Math. Phys.",
    volume = "194",
    pages = "675--705",
    year = "1998"
}

@article{Moore_2000,
	doi = {10.1007/s002200050016},
  
	url = {https://doi.org/10.1007%2Fs002200050016},
  
	year = 2000,
	month = {jan},
  
	publisher = {Springer Science and Business Media {LLC}
},
  
	volume = {209},
  
	number = {1},
  
	pages = {77--95},
  
	author = {Gregory Moore and Nikita Nekrasov and Samson Shatashvili},
  
	title = {D -Particle Bound States and Generalized Instantons},
  
	journal = {Communications in Mathematical Physics}
}

@article{Yi_1997,
    author = "Yi, Piljin",
    title = "{Witten index and threshold bound states of D-branes}",
    eprint = "hep-th/9704098",
    archivePrefix = "arXiv",
    reportNumber = "CU-TP-827",
    doi = "10.1016/S0550-3213(97)00486-0",
    journal = "Nucl. Phys. B",
    volume = "505",
    pages = "307--318",
    year = "1997"
}

@article{Pateloudis_2022,
    author = {Pateloudis, Stratos and Bergner, Georg and Hanada, Masanori and Rinaldi, Enrico and Sch\"afer, Andreas and Vranas, Pavlos and Watanabe, Hiromasa and Bodendorfer, Norbert},
    collaboration = "Monte Carlo String/M-theory (MCSMC)",
    title = "{Precision test of gauge/gravity duality in D0-brane matrix model at low temperature}",
    eprint = "2210.04881",
    archivePrefix = "arXiv",
    primaryClass = "hep-th",
    doi = "10.1007/JHEP03(2023)071",
    journal = "JHEP",
    volume = "03",
    pages = "071",
    year = "2023"
}

@article{Bergner_2021,
    author = {Bergner, Georg and Bodendorfer, Norbert and Hanada, Masanori and Pateloudis, Stratos and Rinaldi, Enrico and Sch\"afer, Andreas and Vranas, Pavlos and Watanabe, Hiromasa},
    collaboration = "Monte Carlo String/M-theory (MCSMC), MCSMC",
    title = "{Confinement/deconfinement transition in the D0-brane matrix model \textemdash{} A signature of M-theory?}",
    eprint = "2110.01312",
    archivePrefix = "arXiv",
    primaryClass = "hep-th",
    reportNumber = "LLNL-JRNL-824792, RIKEN-iTHEMS-Report-21, UTHEP-759, DMUS-MP-21/13",
    doi = "10.1007/JHEP05(2022)096",
    journal = "JHEP",
    volume = "05",
    pages = "096",
    year = "2022"
}

@article{He_2016,
    author = "He, Temple and Mitra, Prahar and Strominger, Andrew",
    title = "{2D Kac-Moody Symmetry of 4D Yang-Mills Theory}",
    eprint = "1503.02663",
    archivePrefix = "arXiv",
    primaryClass = "hep-th",
    doi = "10.1007/JHEP10(2016)137",
    journal = "JHEP",
    volume = "10",
    pages = "137",
    year = "2016"
}

@article{Strominger_2014,
    author = "Strominger, Andrew",
    title = "{Asymptotic Symmetries of Yang-Mills Theory}",
    eprint = "1308.0589",
    archivePrefix = "arXiv",
    primaryClass = "hep-th",
    doi = "10.1007/JHEP07(2014)151",
    journal = "JHEP",
    volume = "07",
    pages = "151",
    year = "2014"
}

@article{Kapec_2017,
    author = "Kapec, Daniel and Lysov, Vyacheslav and Pasterski, Sabrina and Strominger, Andrew",
    title = "{Higher-dimensional supertranslations and Weinberg\textquoteright{}s soft graviton theorem}",
    eprint = "1502.07644",
    archivePrefix = "arXiv",
    primaryClass = "gr-qc",
    reportNumber = "CALT-TH-2015-006",
    doi = "10.4310/AMSA.2017.v2.n1.a2",
    journal = "Ann. Math. Sci. Appl.",
    volume = "02",
    pages = "69--94",
    year = "2017"
}

@article{Kapec_2022,
    author = "Kapec, Daniel and Mitra, Prahar",
    title = "{Shadows and soft exchange in celestial CFT}",
    eprint = "2109.00073",
    archivePrefix = "arXiv",
    primaryClass = "hep-th",
    doi = "10.1103/PhysRevD.105.026009",
    journal = "Phys. Rev. D",
    volume = "105",
    number = "2",
    pages = "026009",
    year = "2022"
}

@article{He_2014,
    author = "He, Temple and Mitra, Prahar and Porfyriadis, Achilleas P. and Strominger, Andrew",
    title = "{New Symmetries of Massless QED}",
    eprint = "1407.3789",
    archivePrefix = "arXiv",
    primaryClass = "hep-th",
    doi = "10.1007/JHEP10(2014)112",
    journal = "JHEP",
    volume = "10",
    pages = "112",
    year = "2014"
}

@article{He_2015,
    author = "He, Temple and Lysov, Vyacheslav and Mitra, Prahar and Strominger, Andrew",
    title = "{BMS supertranslations and Weinberg\textquoteright{}s soft graviton theorem}",
    eprint = "1401.7026",
    archivePrefix = "arXiv",
    primaryClass = "hep-th",
    doi = "10.1007/JHEP05(2015)151",
    journal = "JHEP",
    volume = "05",
    pages = "151",
    year = "2015"
}

@article{Kapec_2014,
    author = "Kapec, Daniel and Lysov, Vyacheslav and Pasterski, Sabrina and Strominger, Andrew",
    title = "{Semiclassical Virasoro symmetry of the quantum gravity $ \mathcal{S}$-matrix}",
    eprint = "1406.3312",
    archivePrefix = "arXiv",
    primaryClass = "hep-th",
    doi = "10.1007/JHEP08(2014)058",
    journal = "JHEP",
    volume = "08",
    pages = "058",
    year = "2014"
}

@article{campiglia2015asymptotic,
    author = "Campiglia, Miguel and Laddha, Alok",
    title = "{Asymptotic symmetries of QED and Weinberg\textquoteright{}s soft photon theorem}",
    eprint = "1505.05346",
    archivePrefix = "arXiv",
    primaryClass = "hep-th",
    doi = "10.1007/JHEP07(2015)115",
    journal = "JHEP",
    volume = "07",
    pages = "115",
    year = "2015"
}

@article{Campiglia_2014,
    author = "Campiglia, Miguel and Laddha, Alok",
    title = "{Asymptotic symmetries and subleading soft graviton theorem}",
    eprint = "1408.2228",
    archivePrefix = "arXiv",
    primaryClass = "hep-th",
    doi = "10.1103/PhysRevD.90.124028",
    journal = "Phys. Rev. D",
    volume = "90",
    number = "12",
    pages = "124028",
    year = "2014"
}

@article{Colferai_2021,
    author = "Colferai, Dimitri and Lionetti, Stefano",
    title = "{Asymptotic symmetries and the subleading soft graviton theorem in higher dimensions}",
    eprint = "2005.03439",
    archivePrefix = "arXiv",
    primaryClass = "hep-th",
    doi = "10.1103/PhysRevD.104.064010",
    journal = "Phys. Rev. D",
    volume = "104",
    number = "6",
    pages = "064010",
    year = "2021"
}

@article{cachazo2014evidence,
    author = "Cachazo, Freddy and Strominger, Andrew",
    title = "{Evidence for a New Soft Graviton Theorem}",
    eprint = "1404.4091",
    archivePrefix = "arXiv",
    primaryClass = "hep-th",
    month = "4",
    year = "2014"
}

@article{Sen_2017,
    author = "Sen, Ashoke",
    title = "{Soft Theorems in Superstring Theory}",
    eprint = "1702.03934",
    archivePrefix = "arXiv",
    primaryClass = "hep-th",
    doi = "10.1007/JHEP06(2017)113",
    journal = "JHEP",
    volume = "06",
    pages = "113",
    year = "2017"
}

@article{Di_Vecchia_2015,
    author = "Di Vecchia, Paolo and Marotta, Raffaele and Mojaza, Matin",
    title = "{Double-soft behavior for scalars and gluons from string theory}",
    eprint = "1507.00938",
    archivePrefix = "arXiv",
    primaryClass = "hep-th",
    reportNumber = "NORDITA-2015-84",
    doi = "10.1007/JHEP12(2015)150",
    journal = "JHEP",
    volume = "12",
    pages = "150",
    year = "2015"
}

@article{Higuchi_2018,
    author = "Higuchi, Sho and Kawai, Hikaru",
    title = "{Universality of soft theorem from locality of soft vertex operators}",
    eprint = "1805.11079",
    archivePrefix = "arXiv",
    primaryClass = "hep-th",
    doi = "10.1016/j.nuclphysb.2018.09.024",
    journal = "Nucl. Phys. B",
    volume = "936",
    pages = "400--447",
    year = "2018"
}

@article{strominger2018lectures,
    author = "Strominger, Andrew",
    title = "{Lectures on the Infrared Structure of Gravity and Gauge Theory}",
    eprint = "1703.05448",
    archivePrefix = "arXiv",
    primaryClass = "hep-th",
    isbn = "978-0-691-17973-5",
    month = "3",
    year = "2017"
}

@article{Bern_2014,
    author = "Bern, Zvi and Davies, Scott and Di Vecchia, Paolo and Nohle, Josh",
    title = "{Low-Energy Behavior of Gluons and Gravitons from Gauge Invariance}",
    eprint = "1406.6987",
    archivePrefix = "arXiv",
    primaryClass = "hep-th",
    reportNumber = "UCLA-14-TEP-104, NORDITA-2014-78",
    doi = "10.1103/PhysRevD.90.084035",
    journal = "Phys. Rev. D",
    volume = "90",
    number = "8",
    pages = "084035",
    year = "2014"
}

@article{Broedel_2014,
    author = "Broedel, Johannes and de Leeuw, Marius and Plefka, Jan and Rosso, Matteo",
    title = "{Constraining subleading soft gluon and graviton theorems}",
    eprint = "1406.6574",
    archivePrefix = "arXiv",
    primaryClass = "hep-th",
    reportNumber = "HU-EP-14-20",
    doi = "10.1103/PhysRevD.90.065024",
    journal = "Phys. Rev. D",
    volume = "90",
    number = "6",
    pages = "065024",
    year = "2014"
}

@article{Bianchi_2015,
    author = "Bianchi, Massimo and He, Song and Huang, Yu-tin and Wen, Congkao",
    title = "{More on Soft Theorems: Trees, Loops and Strings}",
    eprint = "1406.5155",
    archivePrefix = "arXiv",
    primaryClass = "hep-th",
    doi = "10.1103/PhysRevD.92.065022",
    journal = "Phys. Rev. D",
    volume = "92",
    number = "6",
    pages = "065022",
    year = "2015"
}

@article{Maldacena_soft,
    author = "Herderschee, Aidan and Maldacena, Juan",
    title = "{Soft theorems in matrix theory}",
    eprint = "2312.15111",
    archivePrefix = "arXiv",
    primaryClass = "hep-th",
    doi = "10.1007/JHEP11(2024)052",
    journal = "JHEP",
    volume = "11",
    pages = "052",
    year = "2024"
}

@article{Herderschee_2023,
    author = "Herderschee, Aidan and Maldacena, Juan",
    title = "{Three point amplitudes in matrix theory}",
    eprint = "2312.12592",
    archivePrefix = "arXiv",
    primaryClass = "hep-th",
    doi = "10.1088/1751-8121/ad389b",
    journal = "J. Phys. A",
    volume = "57",
    number = "16",
    pages = "165401",
    year = "2024"
}

@article{Lin:2024,
    author = "Lin, Henry W. and Zheng, Zechuan",
    title = "{Bootstrapping ground state correlators in matrix theory. Part I}",
    eprint = "2410.14647",
    archivePrefix = "arXiv",
    primaryClass = "hep-th",
    doi = "10.1007/JHEP01(2025)190",
    journal = "JHEP",
    volume = "01",
    pages = "190",
    year = "2025"
}

@article{Biggs:2025,
    author = "Biggs, Anna and Herderschee, Aidan",
    title = "{Higher-point correlators in the BFSS matrix model}",
    eprint = "2503.14685",
    archivePrefix = "arXiv",
    primaryClass = "hep-th",
    month = "3",
    year = "2025"
}

@article{Donnay:2023,
    author = "Donnay, Laura",
    title = "{Celestial holography: An asymptotic symmetry perspective}",
    eprint = "2310.12922",
    archivePrefix = "arXiv",
    primaryClass = "hep-th",
    doi = "10.1016/j.physrep.2024.04.003",
    journal = "Phys. Rept.",
    volume = "1073",
    pages = "1--41",
    year = "2024"
}

@article{Agrawal:2025,
    author = "Agrawal, Shreyansh and Nguyen, Kevin",
    title = "{Soft theorems and spontaneous symmetry breaking}",
    eprint = "2504.10577",
    archivePrefix = "arXiv",
    primaryClass = "hep-th",
    doi = "10.1103/nbpg-mscq",
    journal = "Phys. Rev. D",
    volume = "112",
    number = "2",
    pages = "L021903",
    year = "2025"
}

@article{McLoughlin_2022,
   title={The SAGEX review on scattering amplitudes Chapter 11: Soft Theorems and Celestial Amplitudes},
   volume={55},
   ISSN={1751-8121},
   url={http://dx.doi.org/10.1088/1751-8121/ac9a40},
   DOI={10.1088/1751-8121/ac9a40},
   number={44},
   journal={Journal of Physics A: Mathematical and Theoretical},
   publisher={IOP Publishing},
   author={McLoughlin, Tristan and Puhm, Andrea and Raclariu, Ana-Maria},
   year={2022},
   month=nov, pages={443012} }

@article{Sethi:1999,
    author = "Sethi, Savdeep and Stern, Mark",
    title = "{Supersymmetry and the Yang-Mills effective action at finite N}",
    eprint = "hep-th/9903049",
    archivePrefix = "arXiv",
    reportNumber = "DUK-CGTP-99-01, IASSNS-HEP-99-17",
    doi = "10.1088/1126-6708/1999/06/004",
    journal = "JHEP",
    volume = "06",
    pages = "004",
    year = "1999"
}
\end{document}